\documentclass[sigconf]{acmart}

\usepackage{tikz}
\usepackage{amsmath}
\usepackage{cancel}
\usepackage[normalem]{ulem}

\usepackage{filecontents}

\usepackage{tikz}
\usepackage{amsmath}

\usepackage{filecontents}
\usepackage{soul}

\usepackage{tikz}
\usepackage{amsmath}
\usepackage{multirow}
\usepackage{booktabs}
\usepackage{MnSymbol}
\usepackage{float}
\usepackage{tabularray,varwidth,enumitem}

\usepackage{caption}
\usepackage{subcaption}

\usepackage{xcolor}
\usepackage{pgfplots}
\pgfplotsset{compat=1.8}
\pgfplotsset{
    width=\textwidth,
}
\usepackage{lipsum}
\usepackage{pgfplotstable}
\usepackage{tabularray}
\usepackage{rotating}

\newcommand{\erik}[1]{\textcolor{cyan}{erik: #1}}

\usepackage[ruled,vlined,linesnumbered]{algorithm2e}
\newcommand{\nosemic}{\renewcommand{\@endalgocfline}{\relax}}% Drop semi-colon ;
\newcommand{\dosemic}{\renewcommand{\@endalgocfline}{\algocf@endline}}% Reinstate semi-colon ;
\let\oldnl\nl% Store \nl in \oldnl
\newcommand{\nonl}{\renewcommand{\nl}{\let\nl\oldnl}}

\SetCommentSty{mycommfont}

\UseTblrLibrary{diagbox}

\usetikzlibrary{pgfplots.groupplots}

\newcommand*\circled[1]{\tikz[baseline=(char.base)]{
            \node[shape=circle,fill,inner sep=1pt] (char) {\textcolor{white}{#1}};}}

\definecolor{codegreen}{rgb}{0,0.6,0}
\definecolor{codegray}{rgb}{0.5,0.5,0.5}
\definecolor{codepurple}{rgb}{0.58,0,0.82}
\definecolor{mGreen}{rgb}{0,0.6,0}
\definecolor{mGray}{rgb}{0.5,0.5,0.5}
\definecolor{mPurple}{rgb}{0.58,0,0.82}
\definecolor{backcolour}{rgb}{0.95,0.95,0.92}

\definecolor{RYB1}{RGB}{80, 99, 42}
\definecolor{RYB2}{RGB}{215, 227, 191}
\definecolor{RYB3}{RGB}{198, 187, 174}
\definecolor{RYB4}{RGB}{146, 205, 220}
\definecolor{RYB5}{RGB}{238, 144, 34}
\definecolor{RYB6}{RGB}{142, 172, 59}

\pgfplotscreateplotcyclelist{colorbrewer-RYB}{
{RYB1!50!black,fill=RYB1},
{RYB2!50!black,fill=RYB2},
{RYB3!50!black,fill=RYB3},
{RYB4!50!black,fill=RYB4},
{RYB5!50!black,fill=RYB5},
{RYB6!50!black,fill=RYB6},
}

\usepackage{lscape} 
\usepackage{pdflscape}
\usepackage{afterpage}
\usepackage{capt-of}
\usepackage{listings}
\usepackage{float}
\usepackage{array,multirow,graphicx}
\usepackage[titletoc]{appendix}
\usepackage{caption}
\usepackage{subcaption}
\usepackage{pifont}% http://ctan.org/pkg/pifont

\definecolor{codegreen}{rgb}{0,0.6,0}
\definecolor{codegray}{rgb}{0.5,0.5,0.5}
\definecolor{codepurple}{rgb}{0.58,0,0.82}
\definecolor{codenavy}{rgb}{0.0,0.2,0.6}
\definecolor{backcolour}{rgb}{0.95,0.95,0.92}

\lstdefinestyle{assembly}{
    backgroundcolor=\color{white}, % Set background color to white
    commentstyle=\color{codegreen},
    keywordstyle=[1]\color{blue},
    keywordstyle=[2]\color{codenavy},
    keywordstyle=[3]\color{codegray},
    numberstyle=\tiny\color{codegray},
    stringstyle=\color{codepurple},
    basicstyle=\ttfamily\small,
    frame=tb, % Add a frame around the code
    framerule=1pt, % Set the width of the frame
    rulecolor=\color{black}, % Set the color of the frame
    breakatwhitespace=false,
    breaklines=true,
    captionpos=b,
    keepspaces=true,
    showspaces=false,
    showstringspaces=false,
    showtabs=false,
    tabsize=4,
    morekeywords={[1]mov,movups},
    morekeywords={[2]rsp,rcx,rax,eax,xmm0},
    morekeywords={[3]QWORD,PTR,DWORD}
}

\definecolor{codegreen}{rgb}{0,0.6,0}
\definecolor{codegray}{rgb}{0.5,0.5,0.5}
\definecolor{codepurple}{rgb}{0.58,0,0.82}
\definecolor{codenavy}{rgb}{0.0,0.2,0.6}
\definecolor{backcolour}{rgb}{0.95,0.95,0.92}

\lstdefinestyle{assembly1}{
    backgroundcolor=\color{white}, % Set background color to white
    commentstyle=\color{codegreen},
    keywordstyle=[1]\color{blue},
    keywordstyle=[2]\color{codenavy},
    keywordstyle=[3]\color{codegray},
    numberstyle=\tiny\color{codegray},
    stringstyle=\color{codepurple},
    basicstyle=\ttfamily\small,
    frame=tb, % Add a frame around the code
    framerule=1pt, % Set the width of the frame
    rulecolor=\color{black}, % Set the color of the frame
    breakatwhitespace=false,
    breaklines=true,
    captionpos=b,
    keepspaces=true,
    showspaces=false,
    showstringspaces=false,
    showtabs=false,
    tabsize=4,
    morekeywords={[1]typedef, struct},
    morekeywords={[2]CHAR, DWORD},
    morekeywords={[3] MAX_PROFILE_LEN , HW_PROFILE_GUIDLEN }
}

\definecolor{ggreen}{HTML}{2CC225}
\definecolor{yyellow}{HTML}{C2C80A}
\definecolor{bbrown}{HTML}{8e4603}
\definecolor{lblue}{HTML}{3333FF}
\definecolor{rred}{HTML}{aa3333}

\newcommand{\csout}[2]{
    \UL\def\temp@sout{\bgroup\markoverwith
        {\textcolor{#1}{\raisebox{1.05ex}{\rule[-0.5ex]{2pt}{1pt}}}}\ULon}
    \temp@sout{#2}
}

\newtheoremstyle{noindent}% <name>
{3pt}% <Space above>
{3pt}% <Space below>
{\upshape}% <Body font>
{}% <Indent amount>
{\itshape}% <Theorem head font>
{.}% <Punctuation after theorem head>
{.5em}% <Space after theorem headi>
{}% <Theorem head spec (can be left empty, meaning `normal')>

\theoremstyle{noindent}
\newtheorem{evasion}{}
\counterwithin{evasion}{section}

\newcommand{\cmark}{\ding{51}}%
\newcommand{\xmark}{\ding{55}}%

\definecolor{codegreen}{rgb}{0,0.6,0}
\definecolor{codegray}{rgb}{0.5,0.5,0.5}
\definecolor{codepurple}{rgb}{0.58,0,0.82}
\definecolor{mGreen}{rgb}{0,0.6,0}
\definecolor{mGray}{rgb}{0.5,0.5,0.5}
\definecolor{mPurple}{rgb}{0.58,0,0.82}
\definecolor{backcolour}{rgb}{0.95,0.95,0.92}

\lstdefinestyle{CStyle}{
    commentstyle=\color{mGreen},
    keywordstyle=\color{magenta},
    numberstyle=\tiny\color{mGray},
    stringstyle=\color{mPurple},
    basicstyle=\sffamily\footnotesize,
    frame=lrtb,
    breakatwhitespace=false,         
    breaklines=true,                 
    captionpos=b,                    
    keepspaces=true,                 
    numbers=left,                    
    numbersep=5pt,                  
    showspaces=false,                
    showstringspaces=false,
    showtabs=false,                  
    tabsize=2,
    language=C
}

\lstdefinestyle{CStyle1}{
    commentstyle=\color{mGreen},
    keywordstyle=\color{magenta},
    numberstyle=\tiny\color{mGray},
    stringstyle=\color{mPurple},
    basicstyle=\sffamily\footnotesize,    frame=lrtb,
    breakatwhitespace=false,         
    breaklines=true,                 
    captionpos=b,                    
    keepspaces=true,                 
    numbers=left,                    
    numbersep=5pt,                  
    showspaces=false,                
    showstringspaces=false,
    showtabs=false,                  
    tabsize=2,
    language=C
}

\lstdefinestyle{mystyle}{
    commentstyle=\color{codegreen},
    keywordstyle=\color{magenta},
    numberstyle=\tiny\color{codegray},
    stringstyle=\color{codepurple},
    basicstyle=\sffamily\footnotesize,
    breakatwhitespace=false,         
    breaklines=true,                 
    captionpos=b,                    
    keepspaces=true,                 
    numbers=left,                    
    numbersep=5pt,                  
    showspaces=false,                
    showstringspaces=false,
    showtabs=false,                  
    tabsize=2,
    language=C
}
\lstdefinestyle{trans}{
    commentstyle=\color{codegray},
    numberstyle=\tiny\color{codegray},
    stringstyle=\color{codepurple},
     basicstyle=\sffamily\footnotesize,
    frame=lrtb,
    breakatwhitespace=false,         
    breaklines=true,                 
    captionpos=b,                    
    keepspaces=true,                 
    numbers=left,                    
    numbersep=5pt,                  
    showspaces=false,                
    showstringspaces=false,
    showtabs=false,                  
    tabsize=2,
     language=[x86masm]Assembler,  escapeinside={\%*}{*)},   
     }     % if you want to add LaTeX within your code

\AtBeginDocument{%
  }

\setcopyright{acmlicensed}
\copyrightyear{2024}
\acmYear{2024}
\acmDOI{XXXXXXX.XXXXXXX}

\acmISBN{978-1-4503-XXXX-X/18/06}

\begin{document}

\title{The Reversing Machine: Reconstructing Memory Assumptions}

%
% The "author" command and its associated commands are used to define
% the authors and their affiliations.
% Of note is the shared affiliation of the first two authors, and the
% "authornote" and "authornotemark" commands
% used to denote shared contribution to the research.
% \affiliation{%

\author{Mohammad Sina Karvandi}
\affiliation{%
  \institution{ Chosun University}
  % \city{Gwangju}
  \country{Republic of Korea}}
\email{karvandi@chosun.kr}

\author{Soroush Meghdadizanjani}
\affiliation{%
  \institution{Stony Brook University}
  % \city{New York}
  \country{USA}}
  \email{smegh@cs.stonybrook.edu}

\author{Sima Arasteh}
\affiliation{%
 \institution{University of Southern California}
 % \city{Los Angeles}
 \country{USA}}
 
 \email{arasteh@usc.edu}

\author{Saleh Khalaj Monfared}
\affiliation{%
 \institution{Worcester Polytechnic Institute}
 % \city{Worcester}
 \country{USA}}
 
 \email{skmonfared@wpi.edu}

 \author{Mohammad K. Fallah}
\affiliation{%
  \institution{ Chosun University}
  % \city{Gwangju}
  \country{Republic of Korea}
  }
 \email{mkfallah@chosun.ac.kr}

 \author{Saeid Gorgin}
\affiliation{%
  \institution{ Chosun University}
    \country{Republic of Korea}}
 \email{gorgin@chosun.ac.kr}

  \author{Jeong-A Lee}
\affiliation{%
  \institution{ Chosun University}
    \country{Republic of Korea}}
 \email{jalee@chosun.ac.kr}

  \author{Erik van der Kouwe}
\affiliation{%
  \institution{Vrije Universiteit Amsterdam}
     \country{Netherlands}} 
 \email{vdkouwe@cs.vu.nl}


\renewcommand{\shortauthors}{Karvandi et al.}

\begin{abstract}
 
Existing anti-malware software and reverse engineering toolkits struggle with stealthy sub-OS rootkits due to limitations of run-time kernel-level monitoring. A malicious kernel-level driver can bypass OS-level anti-virus mechanisms easily. Although static analysis of such malware is possible, obfuscation and packing techniques complicate offline analysis. Moreover, current dynamic analyzers suffer from virtualization performance overhead and create detectable traces that allow modern malware to evade them.

To address these issues, we present \textit{The Reversing Machine} (TRM), a new hypervisor-based memory introspection design for reverse engineering, reconstructing memory offsets, and fingerprinting evasive and obfuscated user-level and kernel-level malware. TRM proposes two novel techniques that enable efficient and transparent analysis of evasive malware: hooking a binary using suspended process creation for hypervisor-based memory introspection, and leveraging Mode-Based Execution Control (MBEC) to detect user/kernel mode transitions and memory access patterns. Unlike existing malware detection environments, TRM can extract full memory traces in user and kernel spaces and hook the entire target memory map to reconstruct arrays, structures within the operating system, and possible rootkits.

We perform TRM-assisted reverse engineering of kernel-level structures and show that it can speed up manual reverse engineering by 75\% on average. We obfuscate known malware with the latest packing tools and successfully perform similarity detection. Furthermore, we demonstrate a real-world attack by deploying a modified rootkit onto a driver that bypasses state-of-the-art security auditing tools. We show that TRM can detect each threat and that, out of 24 state-of-the-art AV solutions, only TRM can detect the most advanced threats.

% We evaluate TRM to reconstruct user-level and kernel-level structures and test them against multiple evasion techniques. We obfuscate known malware with packing tools and successfully perform similarity detection. Furthermore, we demonstrate a real-world attack by deploying a modified rootkit onto a driver that bypasses state-of-the-art security auditing tools. We show that TRM can detect each threat and that, out of 24 state-of-the-art AV solutions, only TRM can detect the most advanced threats.

\end{abstract}

\begin{CCSXML}
<ccs2012>
   <concept>
       <concept_id>10002978.10003006.10003007.10003010</concept_id>
       <concept_desc>Security and privacy~Virtualization and security</concept_desc>
       <concept_significance>500</concept_significance>
       </concept>
   <concept>
       <concept_id>10002978.10002997.10002998</concept_id>
       <concept_desc>Security and privacy~Malware and its mitigation</concept_desc>
       <concept_significance>500</concept_significance>
       </concept>
   <concept>
       <concept_id>10002978.10003022.10003465</concept_id>
       <concept_desc>Security and privacy~Software reverse engineering</concept_desc>
       <concept_significance>500</concept_significance>
       </concept>
   <concept>
       <concept_id>10002978.10003022.10003023</concept_id>
       <concept_desc>Security and privacy~Software security engineering</concept_desc>
       <concept_significance>300</concept_significance>
       </concept>
   <concept>
       <concept_id>10002978.10002991.10002993</concept_id>
       <concept_desc>Security and privacy~Access control</concept_desc>
       <concept_significance>100</concept_significance>
       </concept>
 </ccs2012>
\end{CCSXML}

\ccsdesc[500]{Security and privacy~Virtualization and security}
\ccsdesc[500]{Security and privacy~Malware and its mitigation}
\ccsdesc[500]{Security and privacy~Software reverse engineering}
\ccsdesc[300]{Security and privacy~Software security engineering}
\ccsdesc[100]{Security and privacy~Access control}

\keywords{Reverse-engineering, Binary-analysis, Hypervisor, Malware-analysis, Memory-analysis}

\maketitle

\section{Introduction}

Modern kernel-mode rootkits are a sophisticated class of malware that can potentially reside transparently on a target computer for a long time, evading existing state-of-the-art security measures and anti-virus software. The kernel-level privilege of such malware lets them operate hidden away from the OS memory layout, preventing conventional anti-malware mechanisms from detecting them. 
Early notable examples of such malware include 
 \textit{Stuxnet}~\cite{matrosov2010stuxnet} and \textit{Turla}~\cite{faou2019turla}, which were revealed to have exploited kernel-mode drivers to load their rootkit and carry out hidden execution in kernel mode. Later rootkits have successfully used similar approaches~\cite{kwon2021certified}, including abusing vulnerable drivers~\cite{CVE202017087} and reusing stolen valid driver certificates~\cite{PastDSE,nleak}. Although with the deployment of Kernel Patch Protection (KPP) and Driver Signature Enforcement (DSE), penetrating into the kernel memory became significantly difficult over the last decade, new bypassing mechanisms have been discovered and developed. Such circumvention techniques~\cite{EfiGuard} show that kernel-level integrity can still be compromised. On the other hand, third-party Anti-Virus (AV) and  Endpoint Detection and Response (EDR) software, despite being very effective against malicious user-level applications, suffer from transparency and compatibility issues for patching and monitoring the kernel layout. Additionally, once the rootkit compromises the kernel, even user-mode monitoring can easily be disabled by the rootkits~\cite{positivesec}.

Reverse engineering is critical for blue teams to understand and mitigate modern rootkit threats. For such malware, they typically try to reconstruct the memory layout of binaries to build up profiles and signatures, which can be detected later in other systems~\cite{ligh2014art}. Extracting data structures from the memory dumps and binaries is challenging. Modern malware is highly obfuscated, and much information is lost during compilation, making the profiling process very complicated. Existing reconstruction tools and approaches typically use static analysis~\cite{pagani2021autoprofile,oliveri2022land}, rely on OS-level virtualization or emulations~\cite{slowinska2011howard,lin2010automatic}, and most importantly, are often user-mode supported~\cite{mercier2017dynstruct}, making them futile against kernel-mode rootkits. Due to performance issues and design limitations, dynamic analysis of kernel-mode code is very challenging for automated run-time malware analysis.     
 As the industry gets closer to the adoption of a Zero Trust security model, the principal guideline by Microsoft has been improving internal defense mechanisms such as Virtualization-Based Security (VBS)~\cite{yosifovich2017windows} and Kernel Data Protection (KDP)~\cite{kdp}, as well as signature-based detection and blacklisting stolen certificates and vulnerable drivers~\cite{driverblock,msrc}. As dynamic monitoring of kernel applications comes with its complications and limitations for third-party protection software, anti-virus software often settles for prevention methods based on published malware signatures~\cite{msrc}.

 This trend has pushed researchers and engineers to exercise lower-level security surfaces like hypervisor-based protections~\cite{hypergaurd}. Despite recent efforts \cite{korkin2018divide,tian2019kernel,korkin2017detect} of bare-metal virtualization techniques for malware detection, it is challenging to offer an end-to-end solution with good transparency, performance, and stability. 
 
We resolve these issues by proposing a novel design, incorporating two novel techniques to leverage the hardware-backed virtualization capabilities of modern processors.
% To the best our knowledge, we are the first ones to use Mode-Based Execution Control (MBEC) to efficiently detect mode transitions, which is critical to transparently detect user-kernel interactions.
We designed an end-to-end approach for memory introspection called \textit{The Reversing Machine} (TRM), and implemented it as an open-source framework.
To the best of our knowledge, TRM is the first toolkit capable of tracing and running the kernel dynamically and on bare metal with an acceptable performance penalty. By deploying a custom-made hypervisor, TRM records user and kernel-space memory access transparently in real-time, making it a suitable logging tool against stealthy rootkits.

\paragraph{Contributions} We offer the following contributions:
\begin{itemize}
    \item A novel approach to leverage hypervisor-level hardware features to transparently extract extensive low-level memory traces from kernel-mode malware binaries for signature-based detection, enabling us to counter state-of-the-art in-memory evasion methods in sophisticated malware.

   \item A systematic approach to data structure and calling convention reconstruction from user-mode and kernel-mode binaries. Furthermore, we showcase that simple time series signature matching methods can exploit the extracted logs by TRM for rootkit detection purposes.
    
    \item A prototype implementation of TRM and its extensive evaluation, showcasing the success rate of internal kernel-level structure reconstruction, performance speedup relative to manual reverse engineering, and binary similarity tests for various compilers and commercial packers. Furthermore, we deploy TRM's signature-based detection against multiple real-world user-mode malware and kernel-mode rootkits and compare it to commercial AVs.

    \item A new method of dynamically running and analyzing Windows executables, starting from the entry-point instruction, without the transparency-compromising debug flags \cite{microsoft_process_functions}. In addition, we present a new method to efficiently detect transitions of execution modes.
\end{itemize}
\paragraph{Availability} TRM is open-source and publicly available to foster security research and software engineering: \textbf{the core hypervisor}\footnote{
\small\url{https://github.com/HyperDbg/HyperDbg}
% \small\url{https://anonymous.4open.science/r/TRM-Project}
}, \textbf{structure reconstruction}\footnote{
\small\url{https://github.com/HyperDbg/structure-reconstructor}
% \small\url{https://anonymous.4open.science/r/structure-reconstructor}
}, and \textbf{evaluation results}\footnote{
\small\url{https://github.com/HyperDbg/TRM-results}
% \small\url{https://anonymous.4open.science/r/TRM-results}
}.

% commented out to save space
\begin{comment}
\paragraph{Outline} The remainder of this paper is organized as follows.
In Section~\ref{sec: Technical Background}, we describe a series of technical background concepts. Section~\ref{sec: High-level Overview} provides the high-level design of TRM and its sub-modules. In Section~\ref{sec: Execution Analysis and Control Methods} the detailed execution flow of the proposed analyzer is explained. The reverse engineering process is presented in Section~\ref{sec: Structure Reconstruction}, while Section~\ref{sec: The Memory Analyzer} introduces the memory analyzer of TRM. Section~\ref{sec: Evaluation} provides multiple evaluations. Section~\ref{sec: Applications} describes additional use cases. Section~\ref{sec: Related Works} and Section~\ref{sec: Discussions and Limitations} discuss related works and limitations respectively, and finally Section~\ref{sec: Conclusion} concludes the paper.
\end{comment}

\section{Background}
\label{sec: Technical Background}

% for the BG section, IMO no intro text is needed

\subsection{Hardware virtualization}

A hypervisor or a Virtual Machine Monitor (VMM) configures the CPU to virtually share processors, memory (RAM), and I/O resources~\cite{intel_sdm_vmx} between different virtual machines. Hypervisors are generally categorized into \textit{bare-metal hypervisors (Type-1)} and \textit{OS-hosted hypervisors (Type-2)}. Native hypervisors (Type-1) perform a complete simulation of the hardware, meaning that the VMM is loaded before the operating systems and virtual OSes from the physical host layer are separated. Examples of this category of hypervisors include Xen, VMware ESXi, Microsoft Hyper-V, and KVM.
Non-native (Type-2) hypervisors are executed following the initial boot of the host OS, allowing multiple guest OSes within the same environment \cite{vojnak2019performance} (e.g., VMware Workstation, VirtualBox, Parallels Desktop). Generally, type-1 hypervisors allow for better isolation, scalability, and overall performance. At the same time, running multiple guest OSes on top of an already running OS can provide increased flexibility and ease of use in type-2 hypervisors.

% \paragraph{Intel Virtualization Technology (VT-x)}
{\it Intel's hardware-level virtualization} (VT-x) provides 
hardware-supported infrastructure to facilitate virtualization. VT-x introduces two modes of operation into the x86-64 processors \cite{uhlig2005intel}: root mode and non-root model. \textit{VMX root mode} is used for the VMM itself, allowing it to invoke privileged instructions and configure VMX. Whenever a VM exit occurs in the guest, it is handled in VMX root mode. The guest OS (ring 0) and user-level applications (ring 3), on the other hand, run in \textit{VMX non-root mode}. This mode prevents the execution of privileged VMX instructions, performing a VM exit instead. Thus, the CPU restricts hypervisor-exclusive privileges to the VMX root mode~\cite{takehisa2011aes}.
VMX transitions are controlled by a data structure called \textit{Virtual Machine Control Structure (VMCS)}, managed by the VMM. Each logical processor can have an active VMCS which controls the VMX transitions and configurations.

% \paragraph{Extended Page Tables}
Intel VT-x introduces a hardware-assisted memory extension known as {\it Extended Page Tables} (EPT). EPT is Intel's implementation of Second Level Address Translation (SLAT), automatically translating Guest Physical Addresses (GPA) to Host Physical Addresses (HPA)~\cite{VmwareIntelEptPerformance}.
EPT enhances performance by managing guest page tables in hardware, which was implemented in software in earlier hypervisor implementations~\cite{HardwareAssistedMMU}. Each logical processor can have separate EPT mappings, allowing concurrent access between multiple independent guest operating systems. The VMM can configure and control the EPT paging structures by handling EPT violations, which occur whenever the EPT paging structure entries prevent access to a page~\cite{karvandiHvfs7}.
% It also differs from EPT Misconfiguration which shows that the EPT table is not configured properly. EPT violations only happen when there is no EPT Misconfiguration.

% \paragraph{Mode-Based Execution Control}
% \label{mbec_explanation}
{\it Mode-Based Execution Control} (MBEC) extends EPT to accelerate the execution of unsigned code in guest user-mode while enforcing code integrity in the kernel mode. It was introduced in Intel Kaby Lake, with an equivalent feature present in AMD Zen 2 CPUs~\cite{amd_manual_gmet}. MBEC makes execution permissions depend on whether the address is accessed from kernel-mode or user-mode~\cite{intel_sdm_mbec}.
% In this setup, pages containing codes without signatures can be designated for execution only in the user mode, but remain non-executable in the kernel mode.

% \begin{comment}
% % I removed this part to save space, it does not seem essential to the design
% \subsection{Interrupt-Window Exiting}
% \label{interrupt_window_exiting}
% Interrupt-Window Exiting in VMCS refers to the handling of interrupts in the virtualized environment. VT-x offers special bits in the VMCS, letting the VMM inject interrupts/faults/exceptions to the target core. However, in virtualization, there are situations where the guest machine is not able to process interrupts as the interruptibility state does not allow it.

% In these cases, VMM sets the \textit{Interrupt-Window Exiting} bit in VMCS which leads to a VM exit when the \textit{interrupt window} is open, meaning that the VMM can gain control over injecting interrupts that are directed to the virtual machine. In TRM, this feature is mainly used to inject multiple page faults to the guest.
% \end{comment}

\subsection{EPT Hidden Hooks}
\label{EPThook}
Hidden hooks are breakpoint instructions on a target memory address, where \textit{INT3} instruction is hidden from a VMX non-root observer~\cite{karvandi2022hyperdbg}.
% While the content of memory remained unchanged, the actual execution page contains the breakpoint.
It works by modifying EPT tables such that the page containing breakpoints is executable but not readable. Whenever VMX non-root code tries to read the target page, the hypervisor is notified and it emulates the execution of the instruction with an unmodified version of the target page.

\begin{figure*}[tbp] 
\centering
\includegraphics[width=0.97\linewidth]{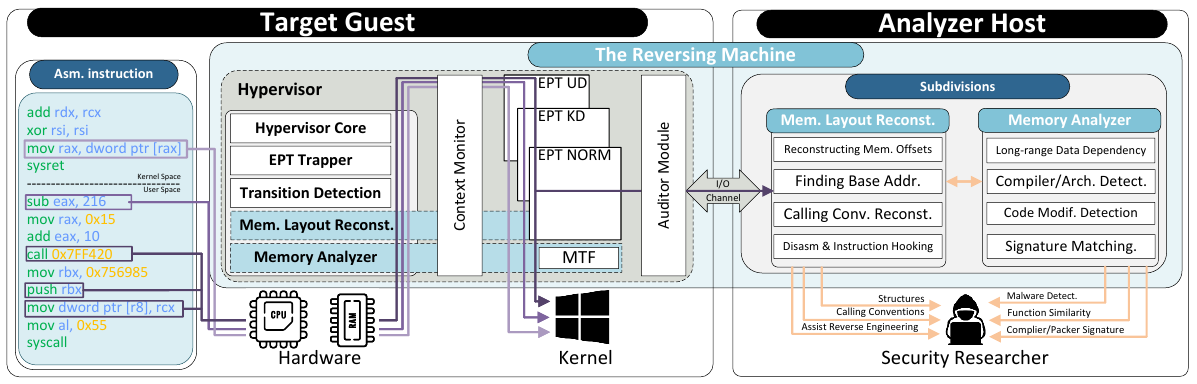}
\vspace{-10pt}
\caption{High-level overview of \textsc{TRM}’s sub-systems and execution flow.}
 \label{fig:TRMArch_and_ExecFlow}
 \vspace{-10pt}
\end{figure*}

% \begin{comment}
% PML never seems to be mentioned in the paper, so no need to introduce it
% \subsection{Page Modification Logging (PML)}
% Intel Page Modification Logging (PML) is a hypervisor feature for tracking accessed virtual machine (VM) memory pages\cite{intel_sdm_pml}. By configuring PML, the system keeps track of dirty pages in a buffer with 512 entries \cite{ren2018phantasy}. %% Given the negligible power consumption overhead associated with PML, TRM effectively minimizes any additional power usage~\cite{bitchebe2020intel} while also reducing live memory migration time.

%%This mechanism is mainly used in TRM to create logs for memory \textit{writes}. However, PML does not record \textit{read} accesses and might give shortcoming results when memory \textit{read} contains useful information. 

% The same behavior of PML can be implemented by using EPT page attributes as it gathers logs from memory \textit{writes}, however, under certain conditions using PML is preferred as it has a significantly lower overhead. This feature is mainly designed to optimize live VM migration.
% \end{comment}

% textit{Other virtualization terminologies and sub-systems used in this paper are explained in Appendix~\ref{appen:A}}.

\subsection{Evasive Malware}
\label{Transparency}
Modern evasive malware is equipped with constantly refined anti-virtualization techniques that allow the malware to detect execution in a virtualized environment, which triggers the malware to exhibit a non-malicious behavior in such circumstances, to increase the difficulty of reverse engineering the malware. Hypervisor-assisted debugging and malware analysis tools aim to counter such efforts by increasing stealth and transparency~\cite{karvandi2022hyperdbg}. We included an overview of modern evasion techniques in Appendix~\ref{appen:B}.

% belows with Overview section, but comes too late if placed there
% \begin{figure*}[tbp] 
% \centering
% \includegraphics[width=0.97\linewidth]{figs/TRM-overview-v5.pdf}
% \caption{High-level overview of \textsc{TRM}’s sub-systems and execution flow.}
%  \label{fig:TRMArch_and_ExecFlow}
% \end{figure*}
%\erik{We need to make sure this figure is consistent with Sections 5-7 in the end}

\section{Threat Model and Goals}
\label{sec:threat_goals}

TRM is designed to analyze state-of-the-art evasive malware. We 
assume the malware runs on Windows, 
although TRM's design could support other guest operating systems with additional effort. We assume that the user (analyst) has privileged access to the host machine to set up TRM, and that the host machine is not compromised. We consider an adversary who can deploy malware equipped with state-of-the-art obfuscation and evasion techniques and can exploit vulnerabilities down to the kernel level (rootkit). Under these conditions, TRM can assist with the following applications:

\noindent\textbf{Reverse-Engineering (G1)} 
TRM assists reverse engineers in assessing their memory assumptions of the binary under analysis by seamlessly and efficiently reconstructing the user-mode and kernel-mode memory layout and recovering structures, calling conventions, and memory offsets.

\noindent\textbf{Malware Analysis (G2)} 
TRM provides malware analysts with behavioral malware analysis based on memory access patterns of the malware to generate high-quality traces of highly privileged malware equipped with the state-of-the-art obfuscation techniques to generate and detect signatures mid-execution, with minimal performance overhead.

\noindent\textbf{Black-box Software Similarity Detection (G3)} In addition to malware analysis scenarios, TRM's memory access pattern signature matching capability offers developers of enterprise and proprietary software the capability to screen for potential unlawful usage of their code by competitors in a black-box manner, even with sophisticated obfuscation and evasion techniques present.

\section{Overview}
\label{sec: High-level Overview}

TRM supports analysts performing reverse engineering and malware analysis on state-of-the-art evasive malware. To use TRM, the analyst starts the hypervisor module, virtualizing an already running system. Each module is integrated with the high-level hypervisor through the log generation mechanism, and is fed with different logs for the occurrence of each event. An event is an incident that is of interest to the module. These logs are then combined to form a structure or make meaningful data, which is reported to the analyst to support their tasks.

Figure~\ref{fig:TRMArch_and_ExecFlow} provides a high-level view of TRM's modules and its data flow. TRM offers both a type-1 hypervisor, including efficient and transparent features for memory interception, and analysis components on the host side of the hypervisor to use the results. The hypervisor-based interception components are connected through an I/O channel to the analysis components on the host. As the processor natively executes instructions, TRM intercepts memory accesses of interest based on several attributes, using hardware support as described in Section~\ref{sec:hypervisor}. The filtered memory accesses are then directed to an auditing module and used to reconstruct low-level characteristics of the code, including entry points, structures, and calling conventions, as described in Section~\ref{sec: Structure Reconstruction}. Building on this information, the memory analyzer module described in Section~\ref{sec: Memory Analyzer} enables higher-level reverse engineering tasks, such as signature extraction for novel binaries based on their memory access traces and dynamic memory signature detection during runtime.

\section{Hypervisor Design}
\label{sec:hypervisor}

In this section, we describe TRM's hypervisor module, presenting novel techniques to transparently capture a target program from the operating system in order to generate a complete memory trace with minimal interference.

%\erik{removed roadmap for now to save space, if we have the space we need to update it due to the changes}
%\textcolor{gray}{
%TRM's innovative methods for transparently monitoring and controlling program execution in a Windows environment are explored. Subsection~\ref{subsection: Finding Entrypoint}, details TRM's hypervisor-based approach for %capturing entry points, while Section~\ref{subsection:Detect Execution in User-mode} discusses TRM's strategies for distinguishing between user and kernel modes. Together, these %insights showcase TRM's effectiveness in enhancing system security with minimal interference.
%}

\subsection{Hypervisor Core}

TRM leverages the hardware-supported virtualization capabilities of modern processors to allow it to monitor all memory accesses occurring in a system from below the kernel. This approach enables the use of high-performance, hardware-facilitated memory tracing via EPT-enabled memory interception capabilities such as EPT Hooks (see Section~\ref{EPThook}), which would otherwise impose a considerable performance penalty. It allows rootkits to be monitored transparently, without leaving monitoring artifacts even for the kernel code. Equipped with EPT-enabled memory interception features, TRM's hypervisor core captures the system's memory traces as it mediates all memory accesses. Specifically, the hypervisor core configures EPT page tables based on Memory Type Range Registers (MTRRs)~\cite{dong2009towards} and emulates memory for interception for physical memory (RAM) ranges. Other memory ranges (such as Memory-Mapped I/O or MMIO) can be directly passed through to the physical hardware. Furthermore, the hypervisor core is also responsible for handling different VMCS configurations and unconditional VM exits~\cite{intel_sdm_unconditional}. The hypervisor core thus sets up an environment to support the monitoring systems described in this section.

\subsection{EPT Trapper}
\label{sec:trapper_ept}

To design an effective memory introspection framework, TRM offers not only comprehensive sub-kernel access to the entire system's memory, but also provides an effective and high-performance means of pruning the trace down to the entries of interest. This poses a challenge, as capturing unfiltered low-level memory traces can yield millions of trace entries in a matter of seconds.
% This is similar to the situation faced in monitoring network packets in high-throughput links, which is addressed by manufacturers in modern networking equipment by specifically designed modules allowing for hardware-supported filtering of packets, a capability absent in general-purpose Intel-based processors.
TRM uses a novel multi-layer EPT layout that directly exploits the hardware-backed virtualization infrastructure of modern processors to address this problem.
Once the hypervisor core is loaded, the EPT Trapper allocates multiple EPT Pointers (EPTPs) with different attributes. This setup allows the VMM to switch seamlessly while maintaining the integrity of EPT entry attributes.
We build on Mode-Based Execution Controls support for user-mode execution prevention. We define the following three main EPTPs used in the trapper component:

\begin{itemize}
  \item User-mode execution disabled: this EPTP contains normal entries while all the user-mode execution bits are disabled. Thus, any user-mode code execution (CPL=3) triggers a VM exit.
  \item Kernel-mode execution disabled: this EPTP contains normal entries while all the kernel-mode (supervisor) execution bits are disabled. Thus, any execution of the kernel-mode code (CPL=0) triggers a VM exit.
  \item Read/Write disabled (execute-only): this EPTP triggers a VM exit whenever read from or written to.
\end{itemize}

The first two EPTPs are used to detect execution in different execution modes (to be discussed in Section~\ref{subsection:Detect Execution in User-mode}), and the last EPTP is used to detect \textit{reads}/\textit{writes} on the memory, for example, to implement EPT Hooks (see Section~\ref{EPThook}). These EPTPs are combined to prune down the trace as depicted in Figure~\ref{fig:TRMArch_and_ExecFlow}. We build on the selective memory access trapping offered by these EPTPs in our analysis modules, including the Memory Layout Reconstruction Module (Section~\ref{sec: Structure Reconstruction}) and Memory Analyzer Module (Section~\ref{sec: Memory Analyzer}).

\begin{comment}
\erik{Removing the log generator for now, if we want to bring it back, we should make clearer how it works and what it contributes}
\subsection{Log Generator}

The core of TRM comes with a Log Generator that can handle and filter thousands of logs simultaneously by employing different hardware features like atomic operations to generate different memory logs related to processes and threads. Different sub-components of TRM then process these logs to extract meaningful information.

The evaluation results are compared in a thread-specific context, while logs are independently handled on individual cores. Each thread generates consecutive results, which might be gathered on different cores. Subsequently, the analyzer module assembles distinct logs from various threads (identified by a unique thread ID) and organizes the actions according to the correct order based on the thread ID.
\end{comment}

\subsection{Detecting Transitions in Execution Modes}
\label{subsection:Detect Execution in User-mode}

TRM allows users to define a custom set of rules based on the user/kernel execution mode of the process for generating logs that can be employed for later analysis. TRM builds on Mode-Based Execution Control (MBEC), delivering substantial performance enhancements compared to native instruction instrumentation approaches such as Qemu \cite{bellard2005qemu}, Bochs \cite{lawton1996bochs}, and DynamoRIO \cite{bruening2012transparent}.
We allocate an additional EPT page table in which user-mode execution is not allowed in any of the corresponding EPT entries (referred to as \textit{MBEC-Denied EPT}). Once the guest OS switches context to the target process, the EPT table will be changed to the \textit{MBEC-Denied EPT} table; thus, no user-mode execution is allowed. Once the application wants to fetch instructions from user-mode, an EPT violation (VM exit) is thrown and the VMM is notified about it. This way, TRM can efficiently intercept user-mode execution.
We also implemented a similar mechanism for backward compatibility with older CPUs, at higher performance overhead but still gaining performance compared to state-of-the-art approaches. We describe this approach in Appendix~\ref{transition_support_for_older_processors}.

\begin{figure}[!t]
\scriptsize
  \centering
  \includegraphics[width=0.85\linewidth]{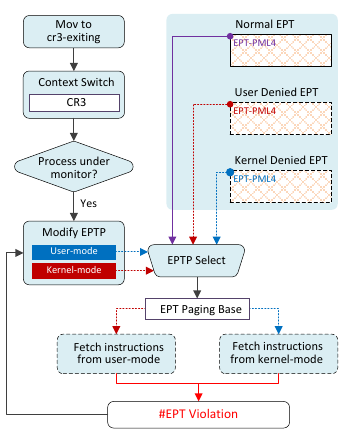}
  \vspace{-10pt}
\caption{The diagram of user-mode and kernel-mode transition detection in TRM.}
  \label{fig:transition}
  \vspace{-15pt}
\end{figure}

\section{Memory Layout Reconstruction}
\label{sec: Structure Reconstruction}

This section presents a methodology to reconstruct information about the memory layout of analyzed malware, including entry point, calling conventions, memory allocations, and memory offsets in structures. These are challenging problems in the context of a hypervisor-based malware analysis tool, which has little information about the memory layout available as it is not integrated with the guest OS. In case of rootkits, this is inherent as we do not trust the guest. We build on the hypervisor-based monitoring features described in Section~\ref{sec:hypervisor} to provide building blocks for the higher-level analysis features presented in Section~\ref{sec: Memory Analyzer}.
% \erik{commented the roadmap for lack of space}
% We first describe actions for different instructions, such as setting EPT hidden hooks and analyzing resulting violations. Subsequently, the focus shifts to the recovery of calling conventions, particularly emphasizing \textit{CALL} instructions and their significance in parameter tracing. Then, we discuss how TRM finds base addresses through memory allocation functions and stack analysis. The section also addresses challenges in the user-mode memory allocations, including page-fault injection and heap manager operations filtering. Finally, we outline a two-phase approach for reconstructing memory offsets, which involves EPT hook placement and finding certain instructions based on disassembly. 

\subsection{Finding Entry Points}
\label{subsection: Finding Entrypoint}
\begin{figure*}[!t] 
\centering
\includegraphics[width=1\linewidth]{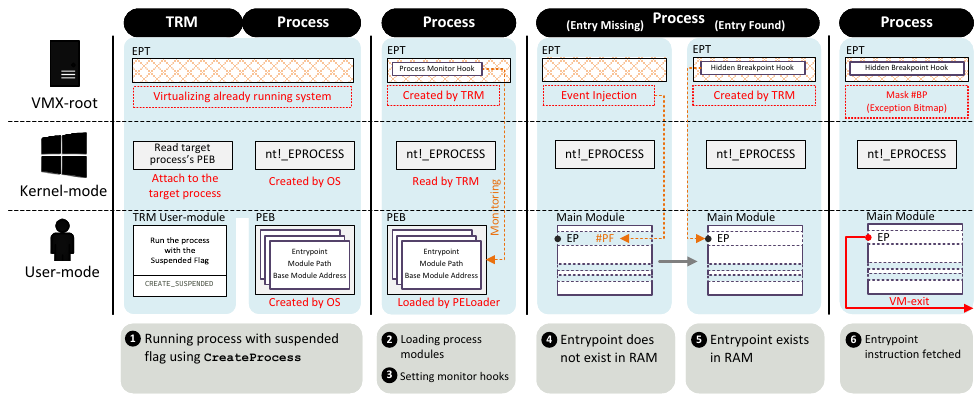}
\vspace{-20pt}
\caption{Intercepting entry points in TRM.}
 \label{fig:TRMEntrypoint_intercept}
\vspace{-10pt}
\end{figure*}

To provide maximal transparency, TRM uses a hypervisor-based approach to find the entry point of an executable file, as shown in Figure~\ref{fig:TRMEntrypoint_intercept}. It avoids the use of debugging flags (\textit{DEBUG\_PROCESS}), which produce distinguishable artifacts, such as debug-optimized heap structures that can reveal the presence of a debugging environment.

In our approach. \circled{1} Windows first loads the executable image with a suspended flag (\textit{CREATE\_SUSPENDED}). \circled{2} Windows then sets a few initial attributes, allocating physical memory, assigning a Page Directory Base Register (\textit{CR3} register on x86) for the base address of the process's page tables, as well as basic process structures, the Process Environment Block (PEB), etc., while allowing it to resume with other processes afterwards. Once Windows switches context to the target process, \circled{3} TRM monitors the PEB to trace any modules loaded by Windows to find the memory address in which the entry point of the program will be loaded. At the point when the target module is loaded and TRM determines its entry point, it checks whether the module is available in the memory or not.
%\textcolor{red}{JUST ONE SUBSECTION! Please remove it!}
%\subsubsection{Break on the entrypoint} 
% \erik{This paragraph does not seem critical, commented it}
% When a shared library (DLL on Windows) is loaded into memory by a process, the OS first checks whether the library is already loaded by another process. If so, Windows only increments its reference count. This allows multiple processes to share the same instance of a library in physical RAM, reducing memory consumption and improving overall system efficiency.
 If the process is already running in the system, then the physical address of the entry point is already available. TRM then revokes the execution permission of the page corresponding to the entry point of the main module, so that an attempt to run it results in an EPT violation and hence a VM exit. Otherwise, due to the lazy loading, Windows only loads the code page into memory after an attempt to run it results in a page fault. However, since the flow of the execution is intercepted using EPT before the page is fetched, \circled{4} TRM injects a page fault to force Windows to make the page contents available. Then, \circled{5} the instructions of the main module are accessible to TRM for logging. At this point, \circled{6} TRM grants the executable permission that it had revoked earlier back to the code page and the execution can be resumed as normal.

\subsection{Disassembly and Instruction Hooking}
\label{Subsec: Different Actions for Different Instructions}

There are different actions corresponding to different instructions used in accessing memory. These actions are determined by employing a disassembler \cite{zydis}. The disassembler is a vital part of TRM and is used in several stages. First, TRM sets EPT hidden hooks (see Section~\ref{EPThook}) in all of the \textit{CALL} instructions, so each time a call instruction is executed, TRM will be notified, and the information provided here is used to feed TRM about the internal state of the program that is under analysis.

 Another situation where the disassembler is used is when an EPT violation occurs. In this case, TRM gathers 16 bytes from the memory of the executing instruction (\textit{GUEST\_RIP}) and examines the instruction that causes the VM exit. This is because the size of the operands of the target instruction that \textit{reads} or \textit{writes} into the memory (e.g., \textit{MOV}) can be determined. This size of operands will later be used to help TRM build each piece of the puzzle that will eventually reconstruct the previously compiled structures.

\subsection{Recovering Calling Conventions}
\label{label:calling_convention}
Identifying \textit{CALL} instructions can be achieved through two approaches. First, TRM could establish EPT hooks on each \textit{CALL} instruction (see Section~\ref{Subsec: Different Actions for Different Instructions}). The second approach uses EPT hooks to monitor stack memory. Execution of \textit{CALL} instructions will cause an EPT violation (VM exit) as the \textit{CALL} instruction pushes the address of the next instruction into the stack. Modifying the stack memory is considered a memory \textit{write}, so it will be handled the same way as simple \textit{MOV} \textit{reads} and \textit{writes}.

\textit{CALL} instructions are essential for TRM in many aspects. First, they can serve as an indicator of a specific condition or status, which is crucial for enhancing pattern recognition algorithms. Second, these instructions are called with different pointers as parameters. For example, once a programmer calls a function, parameters usually contain start addresses of structures and memory locations. Knowing the target environment's calling conventions, we could gather these pointers and enhance the structure reconstruction process. Furthermore, TRM can access all registers once the \textit{CALL} instructions are executed. For example, in a Windows x64 (fastcall) environment, \textit{RCX}, \textit{RDX}, \textit{R8}, and \textit{R9} contain the first four parameters of the function. If there are more parameters to the function, compilers push them to the stack before calling the target function. It is again possible to recover all of them as \textit{PUSH} instructions will be interpreted as a memory modification (\textit{write}) VM exit and will be logged by TRM.
 If several \textit{PUSH} instructions are executed before executing a \textit{CALL} instruction, TRM will conclude that the function contains more than four parameters. For example, in the case of the fastcall calling convention, once the \textit{CALL} instruction is executed, the parameters for 5th, 6th, 7th, and so on are pushed into the stack in \textit{[RSP+20h]}, \textit{[RSP+28h]}, \textit{[RSP+30h]} respectively.
These parameters are then checked with memory validation routines so TRM can infer whether the parameters are valid memory addresses or constant values. If the target value refers to a memory address, TRM will add it to the monitoring list.

TRM's dynamic approach for recovering the calling convention, combined with the data gathered from static analysis tools like IDA Pro, Ghidra, Binary Ninja, etc., is more effective than static approaches alone in determining the target function's calling convention. Static binary analysis tools cannot trace the structure over functions as they might be dynamically loaded through the program. Moreover, TRM helps recover offsets for memory structures, improving the quality of reverse engineering the target binary file.

\subsection{Finding Base Addresses of Memory Allocations}
\label{sec:finding-base}
TRM gathers a list of possible base addresses to recover different structure fields. There are three main sources for finding the base addresses for memory allocations:
invocations of heap allocation functions,
buffers passed as parameters,
and memory references that imply the use of stack buffers.
We will discuss each here.

To detect invocations of heap allocation functions, TRM sets hidden EPT breakpoints on memory allocation functions in both user-mode and kernel-mode (listed in Table~\ref{table:memory_functions}). Using these hooks, TRM will be notified about future memory allocations, the allocated addresses, and the buffer size. Once these functions are called, TRM also retrieves the return address from the stack. Afterwards, TRM can trace the output of the memory allocation (which is the target allocated buffer address).
This approach retrieves all the required information without leaving any visible traces for the guest, preserving transparency.

The second approach uses buffers passed as parameters. Having recovered the calling convention for each function call (as described in Section \ref{label:calling_convention}), TRM lists and stores parameters as these parameters are indicators of base addresses that are passed through functions. For example, the target function has three parameters, while two are valid memory addresses. Although it might not always be true, these valid addresses are likely to be the starting address of previously allocated memory addresses and, thus, are considered base addresses because programmers are most likely to pass the starting base address of structures as the parameter in the corresponding \textit{CALL}s.

Finally, we look for memory references that are indicative of stack-based (local) buffers, which are not traceable by using the first method, while the second will not find all of them. Our third approach is to look for instructions likely to show stack allocations by subtracting from the \textit{RSP} register. Some of these subtractions might be part of the calling convention (Shadow Space), which will be ignored, but in some cases, these subtractions show the memory allocations for local structures in the stack. Another indicator of stack allocation is initialization through \textit{XMM} registers, which compilers often use to optimize the zeroing of stack memory.
% If the compiler attempts to zero the memory using general-purpose registers, it can zero 32-bit (4 bytes) or 64-bit (8 bytes) each time. Instead of using general-purpose registers, compilers prefer to optimize it by zeroing more memory (16 bytes each time) using a single instruction.
Thus, if TRM happens to see the following patterns in program execution, it interprets the \textit{RSP+0xF0} as the target memory address. As there are five consecutive \textit{MOVUPS} instructions, the final size of the buffer will be \textbf{0x50} bytes. Which also might contain some padding bytes to make it aligned).

\noindent\begin{minipage}{\linewidth}
\begin{lstlisting}[style=assembly]
movups [rsp+0xF0],  xmm0
movups [rsp+0x100], xmm0
movups [rsp+0x110], xmm0
movups [rsp+0x120], xmm0
movups [rsp+0x130], xmm0
\end{lstlisting}
\end{minipage}

Using these approaches, we construct a list of allocated buffers within the target program in both user-mode and kernel-mode.

\subsection{User-Mode Memory Allocation and Demand Paging}

Interception of memory allocations made from a user-mode process requires special handling in TRM. Due to demand paging~\cite{kuehner1968demand}, Windows does not allocate (create page table entries) for a user-mode buffer (e.g., from \textit{malloc}) until its first use. 
If TRM tries to access the buffer before Windows allocates physical pages in response to the first page fault, 
% and sets a different memory hook (from the user's \textit{CR3} perspective), 
it will find that the address is not present yet. This behavior applies even when the \textit{MEM\_COMMIT} flag is passed to \textit{NtAllocateVirtualMemory}, the main system-call function for allocating user-mode memory.
To solve this issue, TRM injects a page fault into the target application and implies that the target application tried to access (\textit{read}) the target buffer. When the application continues the user-mode execution, the page fault is delivered and forces the OS to allocate physical memory for the page and set the page table entry.
% After that, the address can be accessed as Windows allocates a valid PFN\erik{what is this?} for the target address and the physical address is now available.
Then, we can use EPT hooks to monitor memory access to the newly allocated addresses~\cite{deng2013spider}.

We inject page faults using Intel's \textit{event injection} mechanism~\cite{intel_sdm_event_injection}. In some cases, the size of the requested buffer is bigger than one single page (4-KB granularity); thus, multiple page faults should be injected to cover the entire address range.
% which is done when the \textit{interrupt-window} is open (see Section~\ref{interrupt_window_exiting}).
Each time that a VM exit occurs,
% (as the result of when the \textit{interrupt-window} is open)
a new page fault will be injected and the VMM repeats these steps until the Windows memory manager brings all of the pages (in the target address range) into the RAM.

Immediately after the application requests a buffer allocation from the operating system through a system call (e.g., \textit{NtAllocateVirtualMemory}), the heap manager of the user-mode application typically tries to access the buffer and perform different operations like dividing the memory into different chunks and adding necessary structures and values before and after the heap (to prevent buffer overflow/underflow and add other protections and security measurements). The heap manager tends to exhibit varying behaviors across different versions of Windows; thus, it is not a good source for investigating memory traces. Nevertheless \textit{reads}/\textit{writes} by the heap manager are also collected by EPT hooks. Still, later, TRM filters these logs mainly based on the instruction pointer (\textit{RIP} register), and those \textit{reads}/\textit{writes} outside the program's main module are not considered for memory trace investigations.

\subsection{Reconstructing Memory Offsets}

\begin{figure*}[!t] 
\centering
\includegraphics[width=0.9\linewidth]{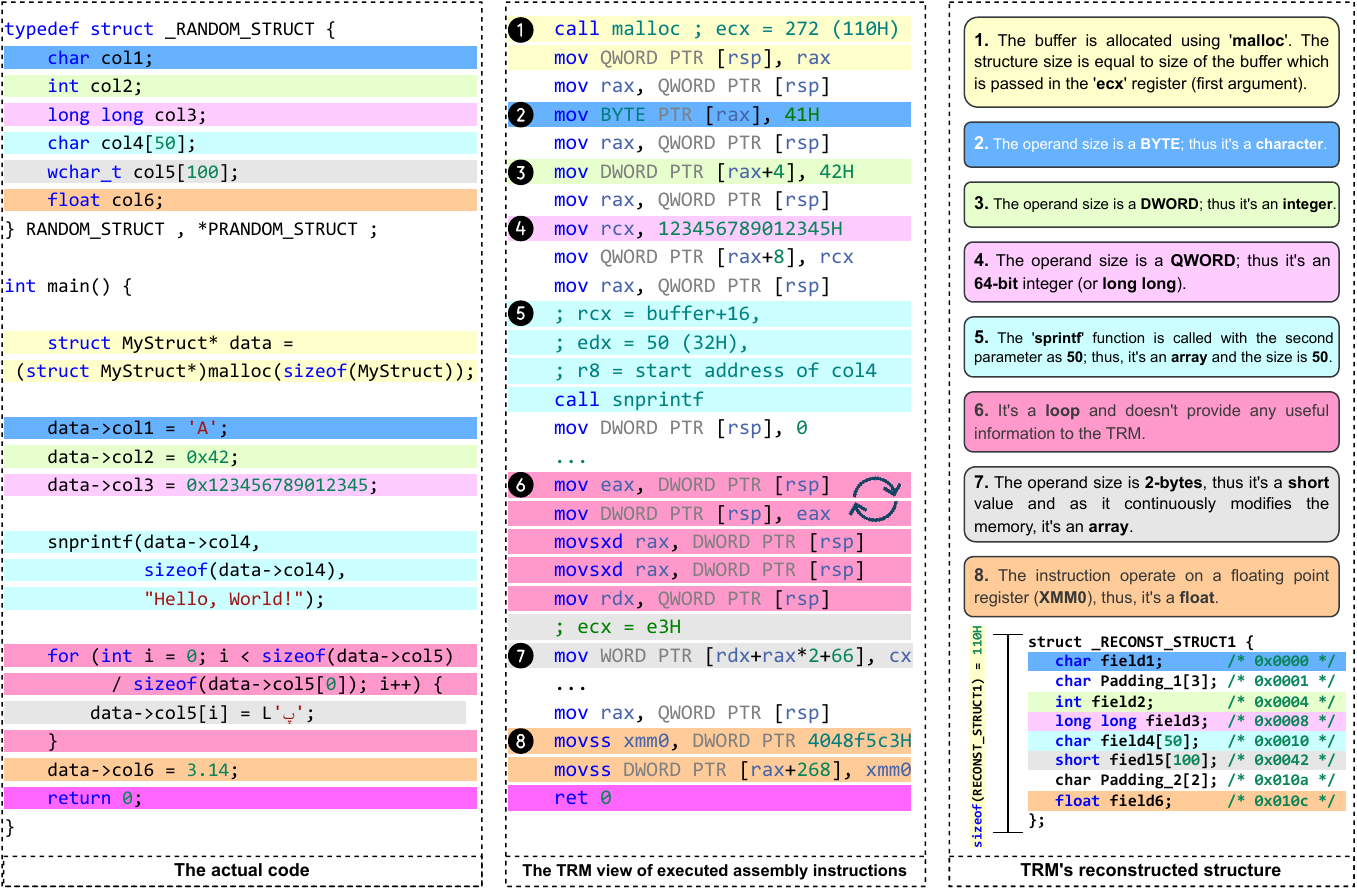}
\vspace{-10pt}
\caption{The process of reconstructing structures.}
\label{overview_reconstructing_struct}
\vspace{-5pt}
\end{figure*}

TRM reconstructs memory offsets in two phases, as shown in Figure~\ref{overview_reconstructing_struct}. The first phase finds the base address of the structure, while the second determines how to interpret the data in it.

To find base addresses, we build on the analysis discussed in Section~\ref{sec:finding-base}. Then, TRM applies monitor EPT hooks to the target memory locations. There might be an initial hint about the structure size, but in some cases, TRM monitors the memory with a default size. After that, the target binary file continues its normal execution; however, as the result of setting EPT monitor hooks~\cite{karvandi2022hyperdbg}, each access (\textit{read}/\textit{write}) will lead to an EPT violation VM exit and as a result, TRM will be notified. Based on the target address, TRM computes the offset (from the base address found earlier) along with the running instruction that caused the EPT violation. These details can be recorded and subsequently examined; alternatively, they may be sent for real-time processing.

In the second phase, the structure reconstruction module reads the previously gathered logs and tries to disassemble instructions to find the necessary details like operand size, whether the instruction performs a floating-point or integer operation (regarding signedness or unsignedness), bitwise operations, array operations, and pointer dereferences.
We continue until all the structure fields are recovered. Note that some fields might not be accessed during the program's lifetime. TRM marks those addresses as arrays of 1-byte characters (\textit{char}).

\begin{comment}
%\erik{commented to save space}
For example, the following instruction is considered as \textit{long long} field.

\begin{lstlisting}[style=assembly]
mov    QWORD PTR [rcx], rax
\end{lstlisting}

While the following instruction is considered as a 32-bit integer.

\begin{lstlisting}[style=assembly]
mov    DWORD PTR [rcx], eax
\end{lstlisting}
\end{comment}

\section{Memory Analyzer}
\label{sec: Memory Analyzer}
This section presents a comprehensive approach to analyze memory structures in binary executables, building on top of the primitives presented in Section~\ref{sec: Structure Reconstruction}. We focus on runtime analysis of long-range data dependencies, finding similarities due to different compilers and architectures, the algorithm used for signature matching, detecting source-code modifications, and combining memory analysis with system and API calls. Each subsection addresses specific challenges and methodologies employed for the memory analysis.

\subsection{Runtime Analysis of Long-Range Data Dependencies}
Finding data dependencies between instructions in binary executable files is a common requirement for binary analysis, including precise call graph construction, malware analysis, exposing hidden behaviors, and binary rewriting. A key challenge in identifying data dependencies is the existence of multiple memory access instructions to the same memory location \cite{zhang2019bda}. Value Set Analysis (VSA), a static binary program analysis technique that tries to identify an approximation of the program state at any given point in the program, is often used to overcome these challenges. VSA is used in well-known static analysis tools like angr \cite{shoshitaishvili2016state} and BAP \cite{brumley2011bap}.
However, while intraprocedural symbolic execution and intraprocedural data flow analysis are used for reverse engineering \cite{jin2014recovering}, it is always challenging to find interprocedural data flow graphs and control flow graphs~\cite{ming2012ibinhunt}.

%The EPT memory access interception approach used in TRM can uncover inter-procedural and intra-procedural Data Dependence Graphs (DDG) dynamically if the memory is accessed (read/write) in the callee function. For implementing this, TRM uses the calling convention details (obtained from Section~\ref{label:calling_convention}) to trace the parameters within functions. The limitation of this method lies in its inability to identify relationships between parameters unless the memory is accessed within the specified callee functions.

TRM's EPT-based memory access interception approach dynamically reveals both interprocedural and intraprocedural Data Dependence Graphs (DDG) whenever memory access occurs within callee functions. To achieve this, TRM leverages calling convention details (as discussed in Section~\ref{label:calling_convention}) to trace function parameters. However, a limitation of this method is its inability to discern relationships between parameters unless the memory is actually accessed within the specified callee functions. However, while the results are not as complete as static analysis, they are much more precise and are helpful in discovering both the interprocedural and intraprocedural DDG.

\subsection{Similarity detection}

\subsubsection{Signature Generation and Matching}
\label{sec:sign_match}

\begin{algorithm}[tbp] 
\small
\caption{Finding the Longest Common Memory Address Pattern by dynamic programming.}
\label{alg:lcmap}
\DontPrintSemicolon
\SetKwInOut{Input}{input}
\SetKwInOut{Output}{output}
\Input{ $P$~, \tcp*{First memory address pattern}  \\
\nonl   $P'$~, \tcp*{Second memory address pattern}
\nonl   $\tau$ \tcp*{Memory alignment threshold}}
\Output{$Result$ \tcp*{LCMAP of the input patterns}}
\SetKwFunction{Zeros}{zeros}
\SetKwFunction{Find}{findLCMAP}
\SetKwFunction{Near}{near}
\SetKwProg{Fn}{function}{}
%\BlankLine
\SetAlgoLined
\Fn{\Find{$P$~,~$P'$~,~$\tau$}}{
	$m~,~n$ $\gets$ {$|P|~,~|P'|$} \tcp*{Sizes of the input patterns}
	$D$ $\gets$ \Zeros{$m$~,~$n$} \tcp*{Initialize $m\times n$ zero matrix}
	\For{$i \gets 1$ to $m$} {
        \For{$j \gets 1$ to $n$} {
			\uIf{\Near{$P_{i-1}$~,~$P'_{j-1}$~,~$\tau$}}{
				$D_{i,j}$ $\gets$ {$D_{i-1,j-1}+1$} \tcp*{Signature matched}
			}
			\Else{
				$D_{i,j}$ $\gets$ {$0$} \tcp*{Signature not matched}
			}			
		}
    }
	$L$ $\gets$ {$max\{D\}$} \tcp*{Length of the LCMAP}
	$I$ $\gets$ {$min\{i~|~D_{i,j}=L\}$} \tcp*{Tail address of the LCMAP}
    $Result$ $\gets$ {$[P_{I-L+1},\ldots,P_{I}]$} \tcp*{The LCMAP}
	\KwRet {$Result$} 
}
\end{algorithm}

TRM adapts the Longest Common Substring dynamic programming algorithm for signature matching, which we will refer to as the Longest Common Memory Address Pattern (LCMAP) algorithm.
We provide our approach in Algorithm~\ref{alg:lcmap}.
The variability introduced by diverse compilers and small changes in the code can lead to minor disparities in the alignment of structure fields. To address this challenge, establishing a threshold, denoted as $\tau$, for alignment becomes necessary. While conducting experiments, we found that setting $\tau=100$ provides a suitable balance, although further optimization may be required to achieve optimal results.

\subsubsection{Alignment of Compiler/Architecture Variances}
%Different compilers result in differences in binaries algorithms and instructions, but the offsets of elements in the data structure remain the same. If the order of accessing memory proves to be the same, it can be concluded that binaries are indeed similar. The same is true even when the same compiler is used to compile the binary in different architectures (e.g., x86 or x64). For example, the default Intel compiler with the default configuration uses \textbf{\textit{vmovdqu}} in x64-bit compiling binaries while the same instruction is not used in its x86 version. However, in both compilers (architectures), the offset of the accessed memory remains the same.

%To measure the similarities, TRM follows two rules. First, it can find the base address of the allocated memory as described in Section~\ref{sec:finding-base} and compute other addresses relatively based on the base address. If it is not possible to find the base address then TRM finds the lowest accessed memory address and measures all the other memory accesses relative to the lowest address.

While different compilers may produce binaries with very different instruction sequences, yet the offsets of elements within data structures typically remain consistent. Such similarities are often dictated by the Application Binary Interface (ABI) to ensure interoperability, with alignment and padding rules often preserving similarities even when there are small changes.
%\erik{Looking at the text again, I don't think we need to make the cross-architecture argument}
% , but in practice this consistency often extends even across different architectures due to functional requirements and similar alignment and padding rules. The main 
% , such as x86 or x64, even when compiled using the same compiler. For instance, while the default Intel compiler utilizes \textit{VMOVDQU} in x64-bit binaries but not in its x86 counterparts, the memory access offsets remain unchanged across both architectures. It is crucial to note that this observation typically holds true for data structures without pointers as the pointer size varies between different 64-bit and 32-bit architectures.
To assess similarities, TRM adheres to two principles. Firstly, it identifies the base address of allocated memory, as detailed in Section~\ref{sec:finding-base}, and computes relative addresses accordingly. Secondly, if the base address cannot be determined, TRM utilizes the lowest accessed memory address as a reference point for measuring other memory accesses.

%For measuring, the similarities, TRM uses an adapted version of the Longest Common Substring (LCS) algorithm.

% leveraging a dynamic programming paradigm to adeptly tackle the challenge by decomposing it into more manageable subproblems and subsequently piecing together the solution.

% \begin{enumerate}
% \item Dynamic Programming Table:
%    - Create a 2D table to store the length of the ACMAPs at each position of the two inputs.

% \item Fill the Table:
%    - Traverse the table and fill it based on the following recurrence relation:
% \begin{scriptsize}
    
% \[ \text{M}[i][j] = 
%     \begin{cases} 
%         0 & \text{if } input_1[i-1] \not \approx input_2[j-1] \\ 
%         \text{M}[i-1][j-1] + 1 & \text{if } input_1[i-1] \approx input_2[j-1] 
%     \end{cases} 
% \]
% \end{scriptsize}

% \item Track Maximum Length:
%    - While filling the table, keep track of the maximum length and its ending position.

% \item Retrieve the Subintervals:
%    - Once the table is filled, retrieve the maximum common interval using the maximum length and its ending position.

% Time Complexity: \(O(m \cdot n)\), where \(m\) and \(n\) are the lengths of the two inputs. This is because we fill a 2D table of size \((m+1) \times (n+1)\).

% Space Complexity: \(O(m \cdot n)\) for the dynamic programming table.

% \end{enumerate}

\subsection{Source Code Modifications}
\begin{figure}[tbp]
    \centering
    \includegraphics[width=1\linewidth]{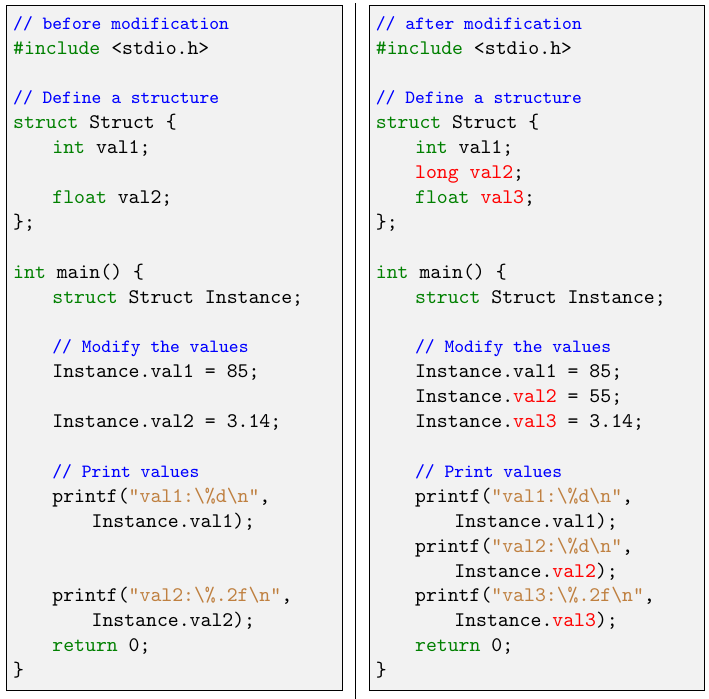}
    \vspace{-20pt}
    \caption{The code before and after modification.}
    \label{fig:code-modification1}
    \vspace{-10pt}
\end{figure}

\begin{figure}[tbp]
    \centering
    \includegraphics[width=1\linewidth]{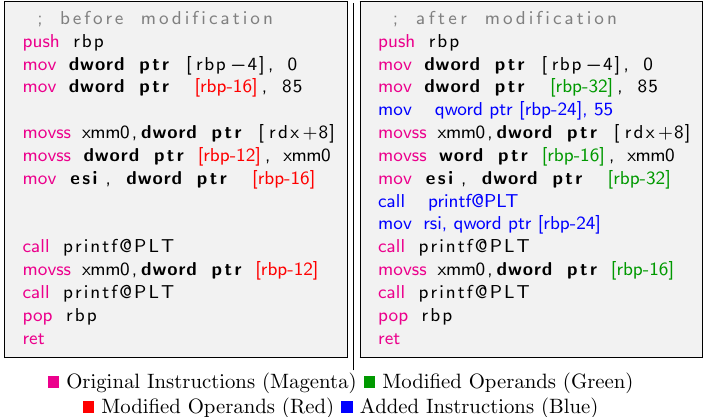}
    \vspace{-20pt}
    \caption{The view of TRM from the modified assembly.}
    \label{fig:code-modification2}
    \vspace{-10pt}
\end{figure}

To detect source-code modifications, TRM initially verifies whether two binaries exhibit similarity (as described in Section~\ref{sec:sign_match}). If similarity is confirmed, TRM proceeds to examine potential code modifications.
Figure~\ref{fig:code-modification1} illustrates a program's source code and its revised version with simple modifications colored in red. These modifications result in the changing of the corresponding assembly codes as shown in Figure~\ref{fig:code-modification2}. In these cases, the instructions in each unmatched subsequent memory pattern are combined, and the comparison within two binaries continues. The result of this comparison shows the differences between the two modified source codes.
Note that we are able to perform this comparison based on runtime traces, so unlike static approaches, we can compare the behavior even for packed and obfuscated malware.

\subsection{Analysis of API Calls and System-calls}

% After gathering different lists of allocated memories along with calling conventions, if the target program tries to use one of these buffers or a special field within a buffer as a parameter to a known API call, or system-call, TRM will flag these fields (or buffers) to a special known structure and further tries to enhance the structure reconstruction for other fields based on the newly discovered structure format. 

% The data for the structures are gathered from Windows WDK, Windows SDK, and other public symbols. Public symbols most of the time come without the definition of functions but can be used to convert the API addresses to the corresponding function names.

Using the previously extracted information, TRM identifies instances where the target program uses buffers or specific buffer fields as parameters for known API calls or system-calls. In such cases, TRM flags these fields or buffers for further analysis, aiming to refine the structure reconstruction process based on the newly unveiled structure format. The structural data is sourced from Windows Driver Kit (WDK), Windows Software Development Kit (SDK), and other public symbols. While public symbols often lack function definitions, they serve as a valuable resource for mapping API addresses to their corresponding function names. This information is critical in understanding malware behavior.

% \section{Memory-based Sandboxing}
% \subsection{Ignoring Memory Writes}

\section{Evaluation}
\label{sec: Evaluation}

In this section, we evaluate TRM's performance with regards to the goals defined in Section~\ref{sec:threat_goals}.
In Section~\ref{sec:struct_reconst} we investigate TRM's ability to reconstruct structures across various execution modes, pertaining to \textbf{G1}. Section~\ref{sec:malware_similarity} explores TRM's ability to generate and detect binaries and malware equipped with sophisticated packing and obfuscation techniques, addressing \textbf{G2} and \textbf{G3}. Finally, Section~\ref{sec:evasive_malware} examines TRM against state-of-the-art evasion techniques, as per \textbf{G2} and \textbf{G3}.

% To evaluate TRM, we assess its ability to reconstruct structures across various execution modes, conducting similarity analyses of malware compiled using different compilers or architectures, and its efficiency in detecting diverse types of malware signatures with different evasion techniques.

\subsection{Data Structure Reconstruction}
\label{sec:struct_reconst}

Here, we focus on exploring how TRM can enhance the automatic reconstruction of structures and aid manual reverse engineering efforts. We gather logs from the top-level hypervisor, and then analyze them later to reconstruct structures, as described in Section~\ref{sec:finding-base}.\\
In this experiment, we focus on recovering Windows kernel data structures. For this purpose, we start the Windows with no application software running on it, and start generating a trace after boot time. The functions shown are exercised by interrupt handlers and user-mode parts of Windows that are scheduled regularly.

\subsubsection{Automatic Reconstruction}
%Table~\ref{table:structure_reconstruction} illustrates various structures situated in different modules/headers that are accessed during different execution modes. Note that the reconstruction percentage depends on the memory accesses (reads/writes) to the target structure (memory), and the matching item is measured based on the reconstruction performance. The completion of the structure reconstruction might be enhanced by dedicating additional time to collecting logs or altering interactions with the system (process), resulting in further modifications to the target process and more structure fields being used.
%We assumed a constant representation for all of the primitive data types as demonstrated in Table~\ref{table:data_type_conversion} in the Appendix.

%This automated process utilizes EPT capabilities to trace execution and log memory accesses, effectively identifying and mapping out data structures without prior knowledge of the software's source code.

Table~\ref{table:structure_reconstruction} provides an overview of various structures accessed across different modules and headers during different execution modes. The percentage of reconstruction completion is contingent upon the frequency of memory accesses, encompassing both \textit{reads} and \textit{writes}, to the target structure. Matching items are assessed based on their reconstruction performance. Enhancing the structure reconstruction process may involve investing additional time in log collection or modifying interactions with the system, thereby inducing further alterations to the target process and facilitating the utilization of more structure fields. We assumed a uniform representation for all primitive data types (for more details, see Table~\ref{table:data_type_conversion} in the Appendix).

\begin{table}[tbp]
\centering
\scriptsize
\caption{Structure Reconstruction Evaluation.}
\label{table:structure_reconstruction}
\vspace{-10pt}
\begin{tblr}{hline{1-2,22,26} = {-}{},
  rowsep=1pt,
}
\textbf{Structure}     & \textbf{Mode}   & \textbf{Module/Header} & \textbf{Rec.\%} & \textbf{Acc.} \\
\_EPROCESS             & K             & NT                     & 29\%               & \cmark             \\
\_ETHREAD              & K             & NT                     & 37\%               & \cmark             \\
\_OBJECT\_HEADER       & K             & NT                     & 100\%              & \cmark             \\
\_KTRAP\_FRAME         & K             & NT                     & 74\%               & \cmark             \\
\_POOL\_HEADER         & K             & NT                     & 80\%               & \cmark             \\
\_LIST\_ENTRY          & U/K           & General/NT             & 100\%              & \cmark             \\
\_DISPATCHER\_HEADER   & K             & NT                     & 68\%               & \cmark             \\
\_RTL\_BITMAP          & K             & win32k                 & 66\%               & \cmark             \\
\_W32PROCESS           & K             & win32k                 & 57\%               & \cmark             \\
\_W32THREAD            & K             & win32k                 & 48\%               & \cmark             \\
\_LARGE\_INTEGER       & U/K           & General/win32k         & 100\%              & \cmark             \\
\_KAPC                 & K             & NT/win32k              & 83\%               & \cmark             \\
{\tiny \_PROCESS\_BASIC\_INFORMATION} & U             & winternl.h             & 100\%              & \cmark             \\
{\tiny \_AVRF\_HANDLE\_OPERATION}     & U             & avrfsdk.h              & 83\%               & \cmark             \\
\_FDICABINETINFO       & U             & fdu.h                  & 100\%              & \cmark             \\
\_OSVERSIONINFOA       & U             & winnt.h                & 100\%              & $\smallstar$             \\
IPAddrEntry            & U             & tcpioctl.h             & 100\%              & \cmark             \\
IPInterfaceInfo        & U             & tcpioctl.h             & 100\%              & \cmark             \\
IPSNMPInfo             & U             & tcpioctl.h             & 69\%               & \cmark             \\
tagHW\_PROFILE\_INFOA  & U             & winbase.h              & 100\%              & $\filledstar$   \\
\SetCell[c=5]{}$\smallstar$ Based on the areas of the modification, TRM was not able to accurately \\ [-1em]
\SetCell[c=5]{} reconstruct this structure. \\
\SetCell[c=5]{}$\filledstar$ This structure uses two consecutive arrays with the same size (CHAR), \\ [-1em]
\SetCell[c=5]{}TRM is not able to distinguish between these two different fields.
\end{tblr}
\vspace{-10pt}
\end{table}

In some scenarios, automatic reconstruction might not be accurate and needs further human interaction to separate results. This especially happens in cases where two arrays in a row with the same data type size are used. For example, the listings below show the definition of one of the failed results. Two consecutive arrays with the same data type (\textit{char}) are defined as demonstrated. There is no way for TRM to separate these two arrays as the gathered memory accesses are consecutive and of the same size. Other information, like the location where the memory is modified (\textit{RIP} register), might be used to separate two different arrays, but that does not necessarily mean there are two arrays (not one array), because the same field of the structure might be accessed somewhere else, it does not imply that the target structure is defined with two fields (arrays) rather than one field (array). As an example, the following code demonstrates a situation where TRM cannot distinguish between two arrays.

\lstdefinestyle{customc}{
  belowcaptionskip=1\baselineskip,
  breaklines=true,
  xleftmargin=\parindent,
  language=C,
  showstringspaces=false,
  basicstyle=\footnotesize\ttfamily,
  keywordstyle=\bfseries\color{green!40!black},
  stringstyle=\color{orange},
}
\begin{lstlisting}[style=assembly1]
typedef struct tagHW_PROFILE_INFOA {
  DWORD dwDockInfo;
  CHAR  szHwProfileGuid[HW_PROFILE_GUIDLEN];
  CHAR  szHwProfileName[MAX_PROFILE_LEN];
} HW_PROFILE_INFOA, *LPHW_PROFILE_INFOA;
\end{lstlisting}

\subsubsection{Comparison with Manual Reverse Engineering}

To assess the effectiveness of TRM compared to manual reverse engineering methods, we conducted a comparative study involving four experienced reverse engineers. To perform an effective comparison, we needed to select a target for the reverse engineering process that would meet the following criteria as closely as possible: 
\begin{itemize}
    \item \textbf{(C1)} Availability of ground-truth (source code or symbols) to assess the validity of the reverse engineering results;
    \item \textbf{(C2)} Kernel-level address space and execution privileges to resemble a rootkit;
    \item \textbf{(C3)} Modules with sufficient complexity to properly represent modern real-world software to allow for meaningful assessment of the performance of the reverse engineering process;
    \item \textbf{(C4)} Reasonable variety and volume of modules to perform a controlled experiment.
\end{itemize}
While an ensemble \textbf{(C4)} of real-world state-of-the-art \textbf{(C3)} rootkits \textbf{(C2)} with leaked source codes \textbf{(C1)} would have been an ideal subject for our reverse engineering study, such an assembly is infeasible to obtain in practice. Therefore, we opted for an alternative that could resemble our ideal scenario as closely as possible. In our experiments, the goal was to reverse and reconstruct memory layout of known Windows kernel-mode structures, particularly those available at the Microsoft Public Symbol servers. 
% \saleh{By selecting known Windows kernel-mode structures, we ensure to evaluate TRM against kernel-level targets (i.e., rootkits) and present our results with a }
We selected 40 structures from the Microsoft Public Symbols of NT module (ntoskrnl) to find the approximate location of where these structures are allocated mainly based on the Windows Research Kernel~\cite{WindowsResearchKernel}. 
% \soroush{todo: add explanation/justification as to why real-world malware was not used instead} \saleh{I added a sentence above}
Then, we placed monitoring hooks on the allocation of these structures for an average duration of approximately one hour. During this monitoring period, we collected data on the structures' field sizes and field locations. This data, along with the reconstructed structures, was then provided to the reverse engineers.
The reverse engineers were divided into two groups, with each group tasked with reversing 20 structures (functionalities). The first group was initially provided with no hints from TRM results and was expected to rely solely on their manual reverse-engineering skills. Once this group concluded their reports, the second group was given the same set of structures but with initial hints derived from the TRM results provided.

Following the completion of their initial tasks, the groups were then switched. The first group, which had previously worked without TRM hints, was given a new set of 20 structures with TRM hints. Meanwhile, the second group, which had initially received TRM hints, was tasked with reversing the same set of structures but this time without any additional assistance from TRM.
% In a comparative study, four experienced reverse engineers, divided into two groups, evaluated TRM against manual methods for reconstructing Windows kernel-mode structures from Microsoft Public Symbol servers. They examined 40 structures from the NT module (ntoskrnl) in two phases, with 20 structures tested in each phase, and outlined the allocation approximation process based on the Windows Research Kernel. Further insights into the methodology of the test, along with a detailed explanation, are provided in Appendix \ref{appen:C3}.
Upon analyzing the results, we observed a significant difference in both the efficiency and the quality of insights between the two groups. On average, the reverse engineering by initial TRM hints for reconstructing the structures leads to $85\%$ and $67\%$ speedups in phases 1 and 2, respectively (see Table \ref{tab:group_phase_timings}).
%were able to speed up their reversing process by 43\%.
% \erik{ideally have a table with this result to make it stand out more, and report data such as the standard error and p-value} 
% \sina{I've added a table to the appendix about the timing but it seems I computed the time incorrectly, I'll ask a friend about the correct way of computing this speedup and will fix it}
Furthermore, the quality and depth of the insights provided by these reverse engineers were substantially better and more accurate (based on the ground-truth structure based on the Microsoft Symbol Server) compared to those who relied solely on manual methods.

\begin{table}[tbp]
\centering
\footnotesize
\caption{Average and Standard Deviation (STD) of Reverse Engineering Time (Hours) Reported by Four Individuals.}
\label{tab:group_phase_timings}
\vspace{-10pt}
\begin{tabular}{lllllll}
\hline
\multirow{2}{*}{\textbf{Phase}} & \multirow{2}{*}{\textbf{\# Struct}} & \multicolumn{2}{c}{\textbf{Manual}} & \multicolumn{2}{c}{\textbf{TRM-Assisted}} & \multirow{2}{*}{\textbf{Speedup}}   \\
  &   & \textbf{AVG Time} & \textbf{STD} & \textbf{AVG Time} & \textbf{STD} &    \\
\hline
1  &  20  &  $15:20'$  &  $0.67$  &  $08:17'$  &  $0.62$  &  $85\%$ \\
2  &  20  &  $17:00'$  &  $0.74$  &  $10:12'$  &  $0.65$  &  $67\%$ \\
\hline
\end{tabular}
\vspace{-10pt}
\end{table}
% This section provides 

% \subsection{Binary Similarity Detection}
\subsection{Malware Similarity Analysis}
\label{sec:malware_similarity}
We use three approaches to evaluate the malware similarity analysis. Firstly, we varied both the compiler and architecture (x86 vs x86\_64) to investigate the potential for detecting these alterations across different binaries and malware samples. Secondly, we applied various obfuscation techniques to the binaries and compared them against state-of-the-art anti-malware tools. Finally, we applied different packers/protectors to the same binary files to assess TRM's ability to identify signatures of packed binary files.

\subsubsection{Compilation Similarity Analysis}

\begin{figure}[tbp]
\scriptsize
  \centering
  \vspace{5pt}
  \includegraphics[width=1\linewidth]{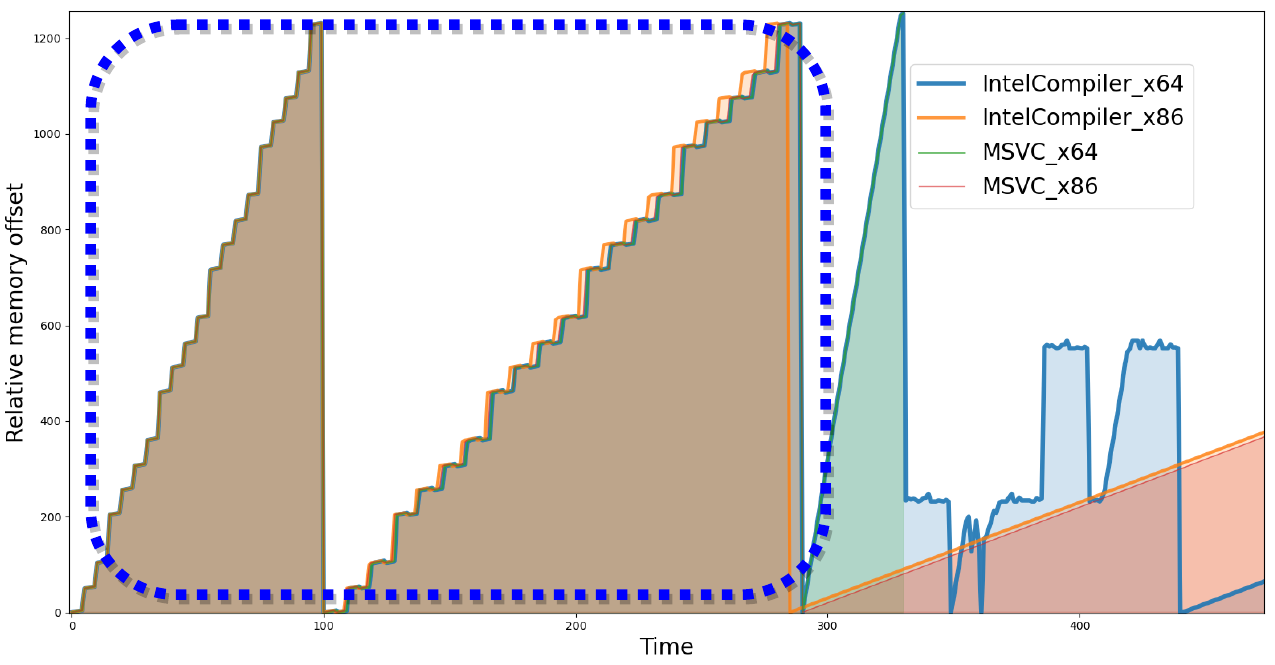}
  \vspace{-20pt}
  \caption{The comparison of accessed offsets for the same code compiled with different compilers/architectures.}
  \label{fig:comparison_compiler_diff}
  \vspace{-10pt}
\end{figure}

For TRM to enable a practical means of defense against malware binaries, it needs to be agnostic towards variance in binaries that is not semantically relevant. To test whether the memory trace of a binary is effectively used for the detection of another obfuscated variation, we introduce differences in architectures and compilers (Intel oneAPI C++ Compiler x86/x64, LLVM-clang x86/x64, Microsoft MSVC x86/x64, PellesC x64, TDM-GCC x86/x64, TinyCC x86/x64). Figure~\ref{fig:comparison_compiler_diff} depicts the relative memory access offsets of a subset of tested compiler/architecture variations of a sample memory-manipulation program across its execution, with the x-axis representing each unique memory operation, showcasing minimal divergence of traces amongst different variations during the memory allocations and accesses of each program (blue window). While small discrepancies emerge due to architectural differences, the traces remain nearly identical, making the traces acquired from a variation viable for use as signatures for defense purposes.
%, as will be explored in later section.

\subsubsection{A Case Study on Signature Matching for Obfuscated Malware}
\label{mimikatz_obfs}
% \textcolor{red}{SOROUSH: PLEASE REWRITE IT}

To compare the effectiveness of TRM against the state-of-the-art anti-malware solutions, we iteratively applied advanced malware obfuscation methods to \textsc{Mimikatz}. \textsc{Mimikatz} is a well-known security tool that can extract plaintext passwords, hash, PIN code, and Kerberos tickets from memory and perform pass-the-hash, pass-the-ticket, or make Golden tickets. There are different methods for detecting usages of this tool \cite{el2020detecting, elgohary2022detecting, smiliotopoulos2022revisiting, matsuda2020detection}, and it is frequently listed on antivirus blacklists. Antivirus programs attempt to prevent its usage due to its common association with red teaming attacks targeting victim machines. Based on testing \textsc{Mimikatz} on \textsc{VirusTotal}, almost all of the anti-malware solutions take this as a virus, therefore, making it the best candidate for testing obfuscation methods. 
% a well-known Windows security tool that is actively used in red teaming attacks
Here, we create five different variations of \textsc{Mimikatz} as follows (see Appendix~\ref{malware_obfs}).
\begin{itemize}
    \item \textbf{OP (Original PE)}: The original executable file without any modification.
    \item \textbf{SP (Shellcode Payload)}: OP is converted into a shellcode using \textit{pe\_to\_shellcode} \cite{pe_2_shellcode} tool and injected like a normal shellcode.
    \item \textbf{ISP (In-memory Shellcode Payload)}: SP is embedded into a new PE file while executed directly on the address space of a container process.
    \item \textbf{RXSP (Remote-memory XOR-encrypted Shellcode Payload)}: SP is XOR encrypted and decrypted in the memory while using remote injection techniques to execute the shellcode into the address space of another process.
    \item \textbf{IXSP (In-memory XOR-encrypted Shellcode Payload)}: SP is XOR encrypted and decrypted in the memory while executed directly on the address space of a container process. 
\end{itemize}

% To show the effectiveness of the method, we used five different versions of \textsc{Mimikatz}. 
The first version represents the original \textsc{Mimikatz} (referred to as OP: Original PE File), for the second version, the original \textsc{Mimikatz} is converted to shellcode by using the \textit{pe\_to\_shellcode} \cite{pe_2_shellcode} tool which converts a PE file to shellcode (referred to as SP: Shellcode Payload), thus it can be then injected like a normal shellcode. As the third sample, the plain shellcode is embedded into a new PE file and it is executed directly on the address space of the container process (referred to as ISP: In-memory Shellcode Payload). In the fourth sample, the shellcode is XORed (encrypted) and decrypted in the memory while it uses remote injection techniques to execute the shellcode into the address space of another process (referred to as RXSP: Remote-memory XOR-encrypted Shellcode Payload). The fifth sample uses the same XOR encryption while it executes the shellcode directly in the address space of the container process (referred to as IXSP: In-memory XOR-encrypted Shellcode Payload).

We compared TRM against the AV solutions reported by \textsc{VirusTotal}, featuring 24 of the best-known ones in Table~\ref{table:anti-virus_tests}. In the first two cases, all AVs detect the signature of \textsc{Mimikatz}. The detection rate drops to $54\%$ when the shellcode is embedded into another executable file. When the shellcode is encrypted, the detection rate drops even further, still some AV solutions detect the remote process injection mechanism and flag the binary file as malicious. 
% For the fifth sample, none of the anti-virus solutions can detect the presence of \textsc{Mimikatz} as it is encrypted and executed directly into the address memory space of the containing process. Even in this case, TRM can detect \textsc{Mimikatz} using its dynamic approach.
For the fifth sample, the \textsc{Mimikatz} was encrypted and executed directly into the address memory space of the containing process. In this case, only TRM detected \textsc{Mimikatz} using its dynamic approach.

As TRM uses a dynamic approach, a sandboxing strategy is needed to minimize the effect of the corresponding malware. After performing a couple of initial memory accesses (based on user configuration), TRM can compare the signature to conclude whether the process is blacklisted or is allowed to be executed. If it is blacklisted, further access to the memory is blocked and the process is terminated. Otherwise, to avoid performance degradation, the memory hooks are removed and the process can continue its normal execution. Usually, a process is not able to perform an effective malicious activity within its first initial accesses however as another defense, TRM blocks and emulates the execution of certain system-calls while implying to the process that the system-call was successfully executed. Though this approach is reasonably effective, certainly it is not the ultimate solution to this problem, here we leave it for future enhancements.

\begin{table}[tbp]
\centering
\scriptsize
\caption{List of Testing Detection of Plain/Encrypted Version of \textsc{Mimikatz} in Different Anti-Malware Solutions.}
\label{table:anti-virus_tests}
\vspace{-10pt}
\begin{tabular}{lccccc
  % colspec={lccccc},
  % rowsep=1pt
}
\hline
              \textbf{AV Solutions} & \textbf{OP} & {\textbf{SP}} & {\textbf{ISP}} & {\textbf{RXSP}} & {\textbf{IXSP}} \\
\hline
\textbf{\textsc{TRM}}   & \cmark & \cmark & \cmark & \cmark & \cmark \\
AhnLab3          & \cmark & \cmark & \xmark & \xmark & \xmark \\
Arcabit                 & \cmark & \cmark & \xmark & \xmark & \xmark \\
Avast                   & \cmark & \cmark & \cmark & \cmark & \xmark \\
AVG                     & \cmark & \cmark & \cmark & \cmark & \xmark \\
Avira (no cloud)        & \cmark & \cmark & \xmark & \xmark & \xmark \\
BitDefender             & \cmark & \cmark & \cmark & \cmark & \xmark \\
ClamAV                  & \cmark & \cmark & \cmark & \xmark & \xmark \\
CrowdStrike Falcon      & \cmark & \cmark & \xmark & \cmark & \xmark \\
Cybereason              & \cmark & \cmark & \cmark & \xmark & \xmark \\
Cylance                 & \cmark & \cmark & \xmark & \cmark & \xmark \\
DrWeb                   & \cmark & \cmark & \cmark & \cmark & \xmark \\
Elastic                 & \cmark & \cmark & \cmark & \cmark & \xmark \\
ESET32         & \cmark & \cmark & \xmark & \xmark & \xmark \\
Fortinet                & \cmark & \cmark & \xmark & \xmark & \xmark \\
Kaspersky               & \cmark & \cmark & \cmark & \xmark & \xmark \\
Malwarebytes            & \cmark & \cmark & \cmark & \xmark & \xmark \\
McAfee                  & \cmark & \cmark & \xmark & \xmark & \xmark \\
Panda                   & \cmark & \cmark & \xmark & \xmark & \xmark \\
SentinelOne (Static ML) & \cmark & \cmark & \cmark & \cmark & \xmark \\
Symantec                & \cmark & \cmark & \cmark & \cmark & \xmark \\
Tencent                 & \cmark & \cmark & \xmark & \xmark & \xmark \\
Trellix (FireEye)       & \cmark & \cmark & \cmark & \cmark & \xmark \\
TrendMicro              & \cmark & \cmark & \xmark & \xmark & \xmark \\
\hline
\end{tabular}
\vspace{-15pt}
\end{table}
% \vspace{-10pt}

\subsubsection{Signature Matching on Packed/Protected Malware}

To examine the resilience of TRM against packing/protecting solutions, we perform obfuscation using 11 state-of-the-art packers/protectors and demonstrate that despite imposition of significant memory overhead to the target binary, both the composition of the structures and the memory accesses sequences of the executables remain intact, rendering TRM effective against these obfuscation methods.  The details of the 11 packer/protector tools are listed in Table~\ref{table:packers_list} (Appendix~\ref{appen:C2})

% As another paradigm, we test whether packer/protector software alters the memory access order or size of the structure fields within the original binary file. According to the tests, despite packers/protectors adding significant memory overhead to the target binary files, they did not alter any structure fields or the sequence of memory accesses to these fields. Consequently, the TRM method remains effective against heavily packed binaries, enabling extraction of the same signature.

% includes a list of packers/protectors that are tested and verified that the memory structures of the packed binaries and the will not be touched.

% \soroush{Brought back table from appendix since it's significant and after removing our comments, we're already below 12 pages}
%\begin{figure*}[!t] 
%\centering
%\captionsetup{justification=centering}
%\includegraphics[width=1\linewidth]{figs/packers.pdf}
%\caption{Similar Malware Detection over different packers using TRM. The horizontal axis represents the original binary of the malware, whereas the Y axis refers to the packed versions of the same malware.}
 %\label{fig:packers}
%\end{figure*}

\subsection{Evasive Malware Detection}
\label{sec:evasive_malware}
The memory signature method used in TRM effectively detects different types of malware signatures that try to evade detection tools using in-memory obfuscation tricks (explained in Appendix~\ref{appen:B}). Following on the case study of \ref{mimikatz_obfs}, we create three different versions of \textsc{Mimikatz} \cite{delpy2014mimikatz} as an executable file, DLL module, and hex assembly code (and loader shellcode) are prepared. 
% \sina{I added another sentence here explaining why we chose Mimikatz, I think it satisfies the above comment}

We equip different variations of \textsc{Mimikatz} with 18 sophisticated evasion techniques (listed in Table~\ref{table:evasive_techniques}) employed by modern malware and use the high-level memory signatures of the buffers of \textsc{Mimikatz}. Using the memory signature matching method described in \ref{sec:sign_match}, TRM is able to successfully detect the signature, subverting all listed evasion methods. Appendix~\ref{appen:B} provides descriptions for some methods, whereas Table~\ref{table:evasive_techniques} presents the sequence of functions for more general techniques.

\begin{table}[tbp]
\centering
\scriptsize
\caption{Evasive Malware Techniques Investigated by TRM.}
\label{table:evasive_techniques}
\vspace{-10pt}
\begin{tblr}{
  colspec={ll},
  rowsep=1pt,
  %vline{2} = {-}{},
  %hline{1-2,14,18} = {-}{},
}
\hline
\textbf{Technique}             & \textbf{Detail Ref.}                      \\ 
\hline
APC Code Injection             & Section~\ref{technique_apc_code_injection}        \\
Early bird APC Code Injection  & Sequence~$1$ (see below)              \\
Process Injection          	   & Sequence~$2$ (see below)   \\
Load PE From Resource          & Sequence~$3$ (see below)   \\
Reflective DLL Injection       & Section~\ref{technique_reflective_DLL_injection}  \\
Module Stomping                & Section~\ref{technique_module_stomping}           \\
Process Hollowing              & Section~\ref{technique_process_hollowing}         \\
Process Doppelgänging          & Section~\ref{technique_process_doppelganging}     \\
Transacted Hollowing           & Section~\ref{technique_transacted_hollowing}      \\
Process Herpaderping           & Section~\ref{technique_process_herpaderping}      \\
Process Ghosting               & Section~\ref{technique_process_ghosting}          \\
Phantom DLL Hollowing          & Section~\ref{technique_phantom_dll_hollowing}     \\
Custom XOR Encoder/Decoder     & \cite{CustomEncDecoder}                   \\
Process Reimaging              & Section~\ref{technique_process_reimaging}         \\
Module Execution Through Fibers          & Sequence~$4$ (see below)     \\
Module Execution Through Thread Pool     & Sequence~$5$ (see below)     \\
Window Hooking                 & Sequence~$6$ (see below)               \\
Map View of Section            & Sequence~$7$ (see below)               \\
\hline
\SetCell[c=2]{}{\scriptsize $1$ $~$ CreateProcessA $\rightarrow$ WriteProcessMemory $\rightarrow$ QueueUserAPC }  \\ [-0.6em]
\SetCell[c=2]{} {\scriptsize $~~~~$ $\rightarrow$ ResumeThread} \\ [-0.2em]
\SetCell[c=2]{}{\scriptsize $2$ $~$ OpenProcess $\rightarrow$ VirtualAllocEx $\rightarrow$ WriteProcessMemory $\rightarrow$ } \\ [-0.6em]
\SetCell[c=2]{} {\scriptsize $~~~~$ CreateRemoteThread, NtCreateThreadEx, or RtlCreateUserThready} \\ [-0.2em]
\SetCell[c=2]{}{\scriptsize $3$ $~$ FindResource $\rightarrow$ SizeofResource $\rightarrow$ LoadResource $\rightarrow$ VirtualAlloc} \\ [-0.3em]
\SetCell[c=2]{}{\scriptsize $4$ $~$ ConvertThreadToFiber $\rightarrow$ VirtualAlloc $\rightarrow$ CreateFiber} \\ [-0.3em]
\SetCell[c=2]{}{\scriptsize $5$ $~$ CreateEvent $\rightarrow$ VirtualAlloc $\rightarrow$ CreateThreadpoolWait} \\ [-0.6em]
\SetCell[c=2]{} {\scriptsize $~~~~$ $\rightarrow$ SetThreadpoolWait} \\ [-0.2em]
\SetCell[c=2]{}{\scriptsize $6$ $~$ LoadLibraryA $\rightarrow$ GetProcAddress $\rightarrow$ SetWindowsHookEx} \\ [-0.3em]
\SetCell[c=2]{}{\scriptsize $7$ $~$ NtCreateSection $\rightarrow$ NtMapViewOfSection $\rightarrow$ RtlCreateUserThread} \\ 
\hline
\end{tblr}
\vspace{-15pt}
\end{table}
%\subsubsection{Rootkits in TRM versus Anti-Virus}
%\textcolor{red}{PLEASE MAKE IT COMPLETE!}
% \input{sec/9-app}

\section{Related Work}
\label{sec: Related Works}
% \erik{We can make this shorter. Let's focus on the most important competition, and use paragraph rather than subsection} 

To the best of the authors' knowledge, TRM is the first framework offering bare-metal simultaneous execution and analysis of kernel-mode and user-mode memory traces with acceptable performance and transparent to malware running in the guest, stemming from the novel techniques enabling TRM to efficiently leverage modern bare-metal virtualization infrastructure. However, TRM is related to the existing literature in several aspects, which we discuss below.

\paragraph{Malware Detection}
Malware analysis solutions can generally be categorized into static and dynamic approaches~\cite{sihwail2018survey}. While static malware analysis might help provide the analyst with a holistic view of the logic of the malware, static approaches have been long shown to struggle with highly-obfuscated malware~\cite{moser2007limits}. Research using dynamic (behavioral) malware analysis has explored a variety of behavioral features such as hash matching~\cite{wicherski2009pehash, botacin2021understanding}, API calls~\cite{sami2010malware,amer2020dynamic,shankarapani2011malware} or specifically system-calls~\cite{canali2012quantitative}, network traffic~\cite{yen2008traffic,rossow2011sandnet}, string pattern matching~\cite{ojugo2019signature, griffin2009automatic}, suspicious activities such as in-memory encryption and mutation~\cite{rad2012camouflage,henson2014memory}, stalling~\cite{kolbitsch2011power}, and anti-VM or anti-debugging~\cite{branco2012scientific}. TRM is similar to approaches that use memory access patterns ~\cite{banin2020detection, banin2016memory,xu2017malware, yucel2020imaging}. Leveraging memory access patterns can be especially effective against fileless malware, which operates without leaving traditional file traces and avoids detection by modifying its signature traits dynamically~\cite{khushali2020review, afreen2020analysis, sudhakar2020emerging, khalid2023insight}. Frequently mutating malware can render hash matching-based, network traffic-based, and string pattern-based methods ineffective~\cite{rhee2011characterizing}, and highly-obfuscated malware equipped with state-of-the-art packing and obfuscation methods that can bypass hash matching-based and anti-VM or anti-debugging detection methods.

\paragraph{Virtualization-Based and Hardware-Assisted Monitoring} TRM uses a hypervisor core to enable sub-kernel monitoring and memory trace capturing. Many hypervisor-assisted frameworks have been deployed for software analysis, testing, kernel debugging, and malware analysis~\cite{dinaburg2008ether,oyama2012detecting, deng2013spider, zhang2018malware, leon2021hypervisor, karvandi2022hyperdbg, fattori2010dynamic} over the last decade.
% Bauman et.al.,\cite{bauman2015survey} present a comprehensive survey on hypervisor-based monitoring, the approaches, applications, and the early tools in this domain.\erik{A 9-year old survey paper does not seem relevant here}
The main advantage of such an approach is the extensive privilege and resource access in sub-kernel execution. This approach has recently been used in high-level software analysis applications including VM guest monitoring~\cite{hsiao2020hardware}, evasive malware analysis~\cite{karvandi2022hyperdbg}, and kernel debugging and monitoring~\cite{ge2020reverse,du2020hart}. Many of these works employ hardware-assisted features such as Intel EPT~\cite{karvandi2022hyperdbg,hsiao2020hardware}, Intel PT~\cite{schumilo2017kafl,schumilo2021nyx}, and ARM Virtualization Extensions (VE)~\cite{nordholz2015xnpro}. Furthermore, there have been several works focusing on hardware-assisted kernel vulnerability detection for AMD64~\cite{pan2017digtool}, ARM~\cite{ning2019understanding},
as well as x86~\cite{jurczyk2018detecting}, using hardware-oriented emulation, taint tracking, and transparent debugging. Despite all these efforts, \textsc{TRM} is the first end-to-end framework to provide a transparent and efficient logging capability based on hypervisor monitoring. TRM adapts processor-enabled features to collect dynamic traces for highly obfuscated malware detection. 

\paragraph{Stripped Metadata Reconstruction} Prior to TRM, \textit{dynStruct} \cite{mercier2017dynstruct} focused on automatic structure reconstruction for user-mode applications using instruction instrumentation (\textit{dynamoRio} \cite{bruening2013building}). Their approach is confined to the user-mode and suffers from significant performance drawbacks, which make it impractical for real-world scenarios due to the slowdown caused by extensive instrumentation. The \textit{OSPREY} framework~\cite{zhang2021osprey} focuses on recovering variables and data structures from stripped binaries. \textit{REWARDS} \cite{lin2010automatic} addresses the reconstruction of data structures from binary code in the user-mode using Intel PIN. \textit{DSCRETE} \cite{saltaformaggio2014dscrete} recovers a variety of the interpretation logic and application data (e.g., images, figures, screenshots). TRM leverages the hypervisor's EPT monitoring capabilities, offering a more efficient and transparent solution suitable for both kernel-mode and user-mode. TRM induces VM exits only for specific accesses to the corresponding structure, avoiding the slowdown associated with full-system or full-application instrumentation, thus providing a faster and more scalable solution.

% \vspace{-15pt}
\section{Discussion and Limitations}
\label{sec: Discussions and Limitations}

\paragraph{Transparency} While TRM cannot claim to guarantee full transparency against evasive malware, it employs various techniques such as the implementation of native hypervisor-level routines to exclude any debugging API in TRM, and against delta timing and side-channel attack defenses, to make itself substantially harder for the anti-debugging, and anti-hypervisor methods to find compared to existing work.

\paragraph{Memory-Oriented Obfuscation} While we demonstrated the effectiveness of TRM against state-of-the-art obfuscation and packing tools, TRM cannot automatically defend against threats where the malicious actor alters the memory layout of the program specifically to circumvent TRM. This requires manual investigation and automation of such defense is pursued as a future work.

\paragraph{Performance} While TRM exhibits minimal performance overhead due to native execution of instructions as opposed to instruction emulation and instrumentation techniques, performance-limiting factors such as VM exits for various EPT violations remain. TRM uses several methods to minimize such violations (VM exits) by applying necessary modifications only to the processes monitored by TRM. For example, the \textit{mov-to-cr3-exiting} feature of VMCS indicates any changes to the \textit{CR3} register, an indication of changes in the processor memory layout that is traditionally a sign of switching context. Although this may result in additional VM exits, it ultimately enhances performance by eliminating the need for EPT violations for each process, as the extra VM exit occurs only during each process context switch and thus, effectively reducing VM exits to one per process transition (from user to kernel and kernel to user) across all processes.
\section{Conclusion}
\label{sec: Conclusion}
%\soroush{added conclusion}

% \textcolor{red}{SOROUSH: PLEASE MAKE IT COMPLETE!!! }
We presented \textit{The Reversing Machine (TRM)}, an end-to-end hypervisor-backed memory introspection design that offers low-level application-targeted and system-wide memory trace capturing, as well as a comprehensive compiler and architecture-neutral memory trace analyzer. TRM employs a novel technique to leverage the virtualization capabilities of modern processors to construct a memory trace capturing mechanism that allows the user to apply custom-defined filters to the memory traces of the system with high performance. We showcased that the calling convention and structure reconstruction capability of TRM, alongside memory trace signature construction and matching, offer a practical and powerful framework for reverse engineering and analysis of highly-obfuscated and highly-privileged malware. We illustrated this by leveraging TRM to reconstruct data structures of various modules in the kernel, as well as investigating the state-of-the-art evasion and obfuscation techniques employed by modern malware. We demonstrated that sophisticated evasion techniques applied to well-known malware can bypass the state-of-the-art modern anti-malware solutions, while displaying the capability of TRM to dynamically intercept, detect, and terminate the execution of malware equipped with state-of-the-art evasion and obfuscation techniques. Overall, TRM offers a comprehensive, hardware-facilitated, and practical memory introspection framework for software security assessment, malware analysis, and black-box software similarity screening based on memory traces.


\newpage 
\bibliographystyle{ACM-Reference-Format}
\bibliography{ref}

\newpage

\section*{Appendices}

\appendix
\label{Appendices}
\section{Technical Details}
\label{appen:A}

This appendix includes detailed technical information that does not fit in the main paper, but has been included here for completeness.

\begin{table}[htbp]
\centering
\caption{Memory Allocation Functions.}
\scriptsize
\label{table:memory_functions}
\begin{tabular}{llc} 
\hline
\textbf{Scope} & \multicolumn{1}{c}{\textbf{Function}}                                                                                                                                                                                                                                                                                     & \textbf{SYSCALL}                                                                                                                                                                                                                                                                                                                   \\ 
\hline
\rotatebox[origin=c]{90}{\textbf{User-mode}} & \begin{tabular}[c]{@{}l@{}}malloc\\calloc\\realloc\\LocalAlloc\\GlobalAlloc\\VirtualAlloc\\CreateFileMapping\\MapViewOfFile\\HeapAlloc\\CoTaskMemAlloc\\NtAllocateVirtualMemory\\NtAllocateVirtualMemoryEx\end{tabular}                                                                                                   & \multicolumn{1}{l}{\begin{tabular}[c]{@{}l@{}}NtAllocateVirtualMemory\\NtAllocateVirtualMemory\\NtAllocateVirtualMemory\\NtAllocateVirtualMemory\\NtAllocateVirtualMemory\\NtAllocateVirtualMemory\\NtCreateSection\\NtMapViewOfSection\\NtAllocateVirtualMemory\\NtAllocateVirtualMemory\\System-call\\System-call\end{tabular}}  \\ 
\hline
              \rotatebox[origin=c]{90}{\textbf{Kernel-mode}} & \begin{tabular}[c]{@{}l@{}}ExAllocatePool\\ExAllocatePoolWithTag\\ExAllocatePoolWithQuota\\MmAllocateContiguousMemory\\MmAllocateNonCachedMemory\\MmAllocatePagesForMdl\\MmAllocatePagesForMdlEx\\MmAllocateSystemMemory\\NtAllocateVirtualMemory\\NtAllocateVirtualMemoryEx\\MmAllocateContiguousNodeMemory\end{tabular} & N/A                                                                                                                                                                                                                                                                                                                                \\
\hline
\end{tabular}
\end{table}
\begin{table}[htbp]
\centering
\caption{Primitive Data-type Conversion/Representation in TRM.}
\footnotesize
\label{table:data_type_conversion}
\begin{tabular}{l|l}
\hline
\textbf{Data type}                                                                                                         & \textbf{Primitive C data types}  \\ 
\hline
int8\_t, BYTE, signed char, bool                                                                                           & char                             \\ 
\hline
unsigned char, uint8\_t                                                                                                    & unsigned char                    \\ 
\hline
\begin{tabular}[c]{@{}l@{}}wchar\_t, int16\_t, WORD, short int, \\signed short, signed short int\end{tabular}              & short                            \\ 
\hline
\begin{tabular}[c]{@{}l@{}}uint16\_t, unsigned short, \\unsigned short int\end{tabular}                                    & unsigned short                   \\ 
\hline
\begin{tabular}[c]{@{}l@{}}int32\_t, DWORD, signed, signed int, \\ long \footnote{The 'long' data-type is defined based on Windows compiler scheme (32-bit).}, long int, signed long, \\ signed long int\end{tabular} & int                              \\ 
\hline
\begin{tabular}[c]{@{}l@{}}uint32\_t, unsigned, unsigned int,\\unsigned long, unsigned long int\end{tabular}               & unsigned int                     \\ 
\hline
\begin{tabular}[c]{@{}l@{}}int64\_t, QWORD, long long int, \\signed long long, signed long long int\end{tabular}           & long long                        \\ 
\hline
\begin{tabular}[c]{@{}l@{}}uint64\_t, unsigned long long,\\unsigned long long int\end{tabular}                             & unsigned long long               \\ 
\hline
float                                                                                                                      & float (32-bit)                   \\ 
\hline
double                                                                                                                     & double (64-bit)                  \\ 
\hline
char*, int*, void* (all pointers)                                                                                          & void* (32/64 bit)                \\
\hline
\end{tabular}
\end{table}

\subsection{Transition Detection Support for Older Processors}
\label{transition_support_for_older_processors}

Since MBEC is a relatively new feature in Intel processors, in order to maintain backward compatibility for older processors (6th generation Skylake and prior processors), TRM design includes a mode switch detection method without requiring the support of MBEC.

\paragraph{\textbf{Regular OS Modification Approach}} In this method, once TRM detects a process context-switch to the target process, the U/S bit of the first regular OS page table (PML4) is set, so the user-mode execution will no longer be allowed. Meanwhile, any page faults will be intercepted through the \textit{Exception Bitmap}, and the virtualized core will continue to reach the user-mode code. Once it needs to fetch user-mode codes, a page fault (\#PF) is thrown which will be intercepted (and ignored) by the VMM. The VMM then detects the user-to-kernel switching and can adjust the OS page tables accordingly. The necessity of handling several indirect page faults in this method can hinder performance and in certain cases negatively impact TRM's transparency.

During its initialization, TRM automatically selects the former or latter approach, based on the support of the processor for Mode-Based Execution Controls, or its lack thereof, respectively. While the second approach equips TRM with backward compatibility with older processors, it imposes extra performance overhead caused by the need to manage several indirect page faults, which may also negatively impact TRM's transparency in certain cases.

\section{Malware Evasion Techniques}
\label{appen:B}

In this appendix, we define several malware evasion techniques discussed in the paper.

% \subsubsection{Process Injection}
\begin{evasion}
\label{technique_process_injection}
{\it Process Injection} refers to a set of techniques \cite{klein2019windows} where malicious codes are injected into the address space of another process and the malicious code executes under the disguise of a legitimate process. This technique is used by malware to avoid detection systems.
\end{evasion}

% \subsubsection{Reflective DLL Injection}
\begin{evasion}
\label{technique_reflective_DLL_injection}
{\it Reflective DLL Injection} uses reflective programming concepts to load a library (DLL) into a host process directly from memory \cite{fewer2008reflective}. The library usually implements a lightweight PE file loader to load itself. This technique allows the library to control its loading and interaction with the host process with minimal access to the host system which facilitates stealthy execution.
\end{evasion}

% \subsubsection{Process Hollowing}
\begin{evasion}
\label{technique_process_hollowing}
{\it Process Hollowing} \cite{leitch2013process} is a method that uses the address memory space of a benign process to replace it with malicious code. In this case, the original process is hollowed out and keeps running under its initial context, allowing the malware to hide its presence. 
\end{evasion}

% \subsubsection{ProcessDoppelg\"{a}nging}
\begin{evasion}
\label{technique_process_doppelganging}
{\it Process Doppelg\"{a}nging} \cite{liberman2017lost} is a code injection technique that utilizes NTFS transactions (known as TxF in Windows \cite{krishna2020comparative}) which is the key functionality of Windows in case of handling files atomicity features. It is used to inject malicious code into the memory section similar to a legitimate process.

Process doppelg\"{a}nging is distinct from process injection and process hollowing as it operates by creating a new process rather than leveraging existing ones. This technique involves impersonating processes without injecting or hollowing, avoiding manipulating other processes' memory or threads. Instead, it exploits filesystem artifacts (e.g., NTFS), making it fileless and leaving minimal traces on the filesystem. 

These attributes allow the process doppelg\"{a}nging to be hidden and resistant to many modern detection and mitigation tools.
\end{evasion}

% \subsubsection{Fileless Malware}
\begin{evasion}
\label{technique_fileless_malware}
{\it Fileless Malware} is a type of malicious program that operates without leaving traditional traces on a computer's file system. Unlike conventional malware that relies on executable files stored on a system, fileless malware exploits legitimate system tools and processes to carry out its malicious activities, making detection and prevention more challenging \cite{sudhakar2020emerging}. These types of malware instead of relying on files often reside in the system's memory. This stealthy approach allows them to evade traditional antivirus and endpoint protection measures, making it a potent threat to computer systems as their footprints are minimized.
\end{evasion}

% \subsubsection{APC Code Injection}
\begin{evasion}
\label{technique_apc_code_injection}
{\it APC Code Injection} An APC (Asynchronous Procedure Call) is a Windows mechanism that runs code asynchronously in the context of a specified thread \cite{ApcInjection}. APC code injection uses this mechanism to deliver and inject codes in the APC queue of the target process.
\end{evasion}

% \subsubsection{Module Stomping}
\begin{evasion}
\label{technique_module_stomping}
{\it Module Stomping}, is done by injecting a benign Windows DLL into a remote process and it is followed by overwriting of the DLL's \textit{AddressOfEntryPoint} with shellcode; subsequently, a new thread is initiated in the target process at the entry point of the benign DLL which contains the embedded shellcode/module~\cite{ModuleStomping}.
\end{evasion}

% \subsubsection{Transacted Hollowing}
\begin{evasion}
\label{technique_transacted_hollowing}
{\it Transacted Hollowing} is a hybrid PE injection technique between process hollowing and process doppelg\"{a}nging which is made by merging elements of these techniques \cite{TransactedHollowing}. The technique involves creating a new process, loading a fresh copy of the NTDLL to avoid detection, and using NTFS transactions to hide the payload. 
\end{evasion}

% \subsubsection{Process Herpaderping}
\begin{evasion}
\label{technique_process_herpaderping}
{\it Process Herpaderping} is a technique that involves altering the content of the executable file on the disk after the image has been mapped, which obscures the true intentions of the process \cite{ProcessHerpaderping}.
\end{evasion}

% \subsubsection{Process Ghosting}
\begin{evasion}
\label{technique_process_ghosting}
{\it Process Ghosting} is a process tampering technique \cite{ProcessGhostingElastic} that is similar to process doppelg\"{a}nging, but uses a delete-pending file instead of a transacted file. Process ghosting involves creating a malicious file, marking it for deletion, copying it into the image segment, suspending the associated handle in the kernel space to remove it from the disk, and then executing the process from a fileless location \cite{kaushik2022implementing}.
\end{evasion}

% \subsubsection{Phantom DLL Hollowing}
\begin{evasion}
\label{technique_phantom_dll_hollowing}
{\it Phantom DLL hollowing} \cite{PhantomDll} searches for a DLL stored on the disk that has not been loaded into memory yet and is sufficiently large to accommodate the malicious payload. Once a suitable DLL is identified, the loader initiates the opening process using transacted NTFS (TxF) \cite{bernardinetti2022pezong}.
\end{evasion}

% \subsubsection{Process Reimaging}
\begin{evasion}
\label{technique_process_reimaging}
{\it Process Reimaging} technique is a post-exploitation evasion method that exploits inconsistencies in how Windows determines process image file locations. It manipulates file object paths and hides the physical location of a malicious process, making it harder to detect and attribute malicious binaries to running processes \cite{ProcessReimagin}. This technique is compared to process hollowing and process doppelg\"{a}nging but is considered easier to execute as it does not require code injection.
\end{evasion}

\section{Evaluation Specifications}

This appendix includes different tables and specifications used in the paper.

\subsection{Malware Obfuscation methods}

\label{malware_obfs}
The following are the hashes of each variation of the malware:

\begin{itemize}
    \scriptsize{
    \item \textbf{OP (Original PE)}: \hfill \\ 61c0810a23580cf492a6ba4f7654566108331e7a4134c968c2d6a05261b2d8a1}
    \item \textbf{SP (Shellcode Payload)}: \hfill \\ 77da1fefb43773e0b09b04e396318b0e93cba13d78c35a2e2e9722975cf190f7
    \item \textbf{ISP (In-memory Shellcode Payload)}: \hfill \\ 0f2ddf65b99af6f9794025d4730b43d613396950de8d0fd235f48a644ca8f5e2
    \item \textbf{RXSP (Remote-memory XOR-encrypted Shellcode Payload)}: \hfill \\ a2e8e18f9a4b6fcf59cac8068e361ca7291fff966fcc413f9aadc71205414846
    \item \textbf{IXSP (In-memory XOR-encrypted Shellcode Payload)}: \hfill \\ 33393cf969e0fbac848d19ae643dcf172a2b77c2679d5b67181ee7da00b476c6
\end{itemize}

Malware obfuscation tests were performed on tests were conducted on February 21, 2024. Future iterations of antivirus solutions may alter their algorithms and flag the files as malicious.

\subsection{Packers/Protectors Solutions}
\label{appen:C2}

\begin{table}[htbp]
\centering
\footnotesize
\caption{Evaluated Packers/Protectors}
\label{table:packers_list}
\begin{tabular}{ll}
\hline
\textbf{Packer/Protector}  & \textbf{Version}  \\
\hline
Themida           & v3.1.8.0 \\
VMProtect         & v3.8.6   \\
ASPack            & v2.43    \\
UPX               & v4.2.2   \\
Engima            & v7.40    \\
Obsidium x64      & v1.8.2.9 \\
PELock            & v2.11    \\
TELock            & v0.98    \\
Yoda              & v1.3     \\
PECompact         & v2       \\
Petite compressor & v2.4     \\
\hline
\end{tabular}
\end{table}

% \usepackage{tabularray}

% \subsection{The Timing Spent of Reverse Engineering Structures}
% \label{appen:C3}
% \begin{table}[htbp]
% \centering
% \footnotesize
% \caption{Detailed Reverse Engineering Timings by Group and Phase}
% \label{tab:group_phase_timings}
% \begin{tabular}{lllll}
% \hline
% \textbf{Group} & \textbf{Phase} & \textbf{Method}       & \textbf{Structure Count} & \textbf{Average Time (hours)}   \\
% \hline
% 1     & 1     & Manual       & 20              & $\simeq15:20$ \\
% 2     & 1     & TRM-Assisted & 20              & $\simeq08:17$  \\
% 1     & 2     & Manual       & 20              & $\simeq17:00$ \\
% 2     & 2     & TRM-Assisted & 20              & $\simeq10:12$ \\
% \hline
% \end{tabular}
% \end{table}

% \begin{table}[htbp]
% \centering
% \footnotesize
% \caption{Average and Standard Deviation (STD) of Reverse Engineering Time (Hours) Reported by Four Individuals.}
% \label{tab:group_phase_timings}
% \begin{tabular}{lllllll}
% \hline
% \multirow{2}{*}{\textbf{Phase}} & \multirow{2}{*}{\textbf{\# Struct}} & \multicolumn{2}{c}{\textbf{Manual}} & \multicolumn{2}{c}{\textbf{TRM-Assisted}} & \multirow{2}{*}{\textbf{Speedup}}   \\
%   &   & \textbf{AVG Time} & \textbf{STD} & \textbf{AVG Time} & \textbf{STD} &    \\
% \hline
% 1  &  20  &  $15:20'$  &  $0.67$  &  $08:17'$  &  $0.62$  &  1.85 X \\
% 2  &  20  &  $17:00'$  &  $0.74$  &  $10:12'$  &  $0.65$  &  1.67 X \\
% \hline
% \end{tabular}
% \end{table}

\end{document}